\documentclass{aa}
\usepackage{graphicx}
\usepackage{float}
\graphicspath{ {./plots/} }
\usepackage{txfonts}
\usepackage{natbib}
\bibpunct{(}{)}{;}{a}{}{,}
\usepackage{multicol}
\usepackage{array,multirow}
\usepackage{footnote}
\usepackage{url}
\usepackage[titletoc]{appendix}
\usepackage[colorlinks=true,
    linkcolor=blue,   
    citecolor=blue,   
    urlcolor=blue]{hyperref}
\usepackage{cleveref}
\usepackage{soul}
\usepackage{booktabs}
\usepackage{tabularx}
\usepackage{subcaption}
\usepackage{orcidlink}
\usepackage{placeins}

\usepackage[switch]{lineno}

\usepackage{color} 
\usepackage[dvipsnames]{xcolor}

\begin{document}

   \title{Catching the 2021 $\gamma$-ray flare in the blazar TXS~2013$+$370 }

\author{G.~Michailidis\inst{1} \orcidlink{0009-0004-6943-1658}
E.~Traianou \inst{2,3} \orcidlink{0000-0002-1209-6500}
\and N.~Marchili\inst{4} \orcidlink{0000-0002-5523-7588}
\and G.~F.~Paraschos\inst{3} \orcidlink{0000-0001-6757-3098}
\and T.~P.~Krichbaum\inst{3} \orcidlink{0000-0002-4892-9586}
\and U.~Bach\inst{3} \orcidlink{0000-0002-7722-8412}
\and P.~A.~Vergara~de~la~Parra\inst{5}
\and Dong-Jin Kim\inst{7} \orcidlink{0000-0002-7038-2118}
\and V. M. Pati\~no-\'Alvarez \inst{3,9} \orcidlink{0000-0002-5442-818X}
\and M.~Kadler\inst{8} \orcidlink{0000-0001-5606-6154}
\and M.~A.~Gurwell\inst{6} \orcidlink{0000-0003-0685-3621}
}

\institute{Department of Physics, Aristotle University of Thessaloniki, 54124 Thessaloniki, Greece
\and Interdisziplin\"ares Zentrum f\"ur Wissenschaftliches Rechnen (IWR), Universit\"at Heidelberg, Im Neuenheimer Feld 205, 69120 Heidelberg, Germany
\and Max-Planck-Institut f\"ur Radioastronomie, Auf dem H\"ugel 69, D-53121 Bonn, Germany
\and INAF -- Istituto di Radioastronomia, Via P. Gobetti 101, 40129 Bologna, Italy
\and Owens Valley Radio Observatory, California Institute of Technology, Pasadena, CA 91125, USA
\and Center for Astrophysics | Harvard \& Smithsonian, 60 Garden Street, Cambridge, MA 02138, USA
\and CSIRO, Space and Astronomy, PO Box 76, Epping, NSW 1710, Australia
\and Institut für Theoretische Physik und Astrophysik, Universität Würzburg, Emil-Fischer-Str. 31, 97074 Würzburg, Germany
\and Instituto Nacional de Astrof\'isica, \'Optica y Electr\'onica, Luis Enrique Erro $\#1$, Tonantzintla, Puebla 72840, M\'exico
}

   \date{Received ????? / Accepted ?????}

  \abstract{The $\gamma$-ray–loud blazar \object{TXS~2013+370}, a powerful multiwavelength emitter at $z=0.859$, underwent an exceptional GeV outburst in late 2020–early 2021. In this work, we present full-polarization VLBI imaging at 22, 43, and 86\,GHz (2021 February 11) together with contemporaneous single-dish monitoring (OVRO 15\,GHz; SMA 226\,GHz) and \textit{Fermi}–LAT light curves, to localize the high-energy dissipation site and probe the inner-jet magnetic field. The images enabled us to study the jet structure and field topology on sub-parsec scales, revealing a compact near-core knot at $r\simeq40$–$60~\mu$as contemporaneously with the GeV flare and a flat, core-dominated spectrum ($\alpha\gtrsim-0.5$). Core has strong linear polarization and exhibits a $\sim50^{\circ}$ EVPA rotation at 86\,GHz; pixel-based and integrated fits yield a high, uniform rotation measure, $\mathrm{RM}=(7.8\pm0.2)\times10^{4}\ \mathrm{rad\,m^{-2}}$, consistent with an external Faraday screen. Cross-correlation of \textit{Fermi}–LAT and 15\,GHz light curves shows a highly significant peak with the $\gamma$ rays leading by $\Delta t=(102\pm12)$\,d; adopting $\beta_{\rm app}=4.2\pm0.5$ and $\theta=4.1^{\circ}\!\pm0.2^{\circ}$ implies a de-projected separation $\Delta r_{\gamma-15}=(2.71\pm0.47)$\,pc and locates the GeV emission between the jet apex and $\sim0.42$\,pc (in the $1\sigma$ range) downstream. Our results do not uniquely pinpoint the emission site; rather, they support two valid scenarios. The $\gamma$-ray production occurs within-BLR ($\sim0.07$\,pc), where external-Compton scatters to $\gamma$-rays optical and UV photons, and beyond-BLR reaching $\sim0.42$ pc (1$\sigma$) within the inner parsecs, where external-Compton scattering of dusty-torus infrared photons dominates. Both scenarios are compatible in the allowed range of emission distances while opacity-driven core shifts modulate the observed radio–$\gamma$ delay without requiring large relocations of the dissipation zone.

}


\titlerunning{TXS~2013$+$370: A $\gamma$-ray loud blazar at ultra high angular resolution}
\authorrunning{G. Michailidis et.al} 

\maketitle

\section{Introduction}
\label{sec:intro}

Blazars are the most extreme manifestation of Active Galactic Nuclei (AGN). These objects feature relativistic plasma jets launched from supermassive black holes that are oriented close to our line of sight, and emitting violently across the electromagnetic spectrum \citep[e.g.,][]{1995PASP..107..803U, Ulrich_1997, Sikora_2001, Cavaliere_2002}. A key question that concerns the physical processes driving the $\gamma$-ray emission in jets is the location of its production site. Observational findings and theoretical models suggest that high-energy emission can originate both in regions close to the central engine and further downstream in the jet \citep[e.g.,][and references therein]{2001ApJ...556..738J, 2014A&A...571L...2R,  2019AA...630A..56P, 2020ApJ...891...68C}. To understand the physical mechanisms behind these emission processes, leptonic models provide a comprehensive theoretical framework. In this framework, the photons that are up-scattered by the relativistic leptons to $\gamma$-ray energies, are either the same synchrotron photons radiated by the jet (Synchrotron-Self-Compton, SSC; \citealt{1992ApJ...397L...5M}) or photons from the jet surroundings (External Compton, EC; e.g., \cite{1994ApJ...421..153S, 2008MNRAS.386L..28G,2009ApJ...692...32D}). With increasing distance from the black hole, possible reservoirs of these seed photons include the accretion disk \citep{1992A&A...256L..27D}, the broad-line region (BLR e.g., \cite{1993ApJ...416..458D,2012ApJ...758L..15D}), the dusty torus \citep{2000ApJ...545..107B,1999ApJ...514..138K}, or even cosmic microwave background photons \citep{2008MNRAS.385..283C}.

Millimeter Very Long Baseline Interferometry (mm-VLBI) observations are particularly suited for investigating the $\gamma$-ray emission origin in blazars, especially when they are combined with broadband variability monitoring \citep[e.g.,][]{2017A&ARv..25....4B,2023A&A...669A..32P}. While $\gamma$-ray detectors have poor angular resolution ($\sim$0.1$^{\circ}$ for \textit{Fermi}-LAT at 1 GeV; \citealt{2009ApJ...697.1071A}), mm-VLBI provides extremely high resolution views of innermost jet regions, unaffected by synchrotron opacity. Moreover, polarimetric VLBI observations probe the magnetic field configuration in relativistic jets, which is also crucial for particle acceleration (\cite{paraschosa}; \cite{paraschosb}). MHD simulations show that as jets propagate, plasma instabilities develop toroidal components that aid collimation and stability \citep{McKinney_2009,Lyutikov_2013}, directly influencing particle acceleration through reconnection and shock processes \citep{Sironi_2014,Sironi_2016}.

\indent The compact radio source \object{TXS~2013+370} is a powerful $\gamma$-ray loud blazar located at redshift $z = 0.859$ \citep{2013ApJ...764..135S}, and hosts a supermassive black hole with a mass of hosts a supermassive black hole with a mass of $(4^{+8.6}_{-2.7}) \times 10^{8}$\,M$_\sun$ \citep{2015MNRAS.448.1060G}, where the uncertainty reflects the ~0.5-0.6 dex systematic error inherent to virial mass estimates \citep{2006ApJ...641..689V}. Previous studies \citep{2000ApJ...542..740M,2001ApJ...551.1016H,2012ApJ...746..159K} established the firm association between \object{TXS~2013+370} and a \textit{Fermi}-Large Area Telescope (LAT) source through correlated radio and $\gamma$-ray variability, revealing a two-component spectral energy distribution best described by EC processes involving seed photons from a dusty torus environment. Subsequent VLBI studies by \citet{refId0} constrained the $\gamma$-ray emission location to approximately 1~pc from the jet apex and revealed a transition from parabolic to conical jet expansion at parsec scales.\\
In this study, we investigate the morphological evolution, magnetic field structure, and spectral properties of the radio jet during an exceptional $\gamma$-ray flaring episode that began in December 2020. We conducted full polarimetric Target of Opportunity (ToO) observations using the Very Long Baseline Array (VLBA) + Effelsberg at 22, 43\,GHz, complemented by VLBA observations at 86\,GHz, achieving angular resolutions down to $\sim$0.14mas ($\sim$1.1\,pc). By combining VLBI imaging with kinematic analysis and correlated flux density variability across radio, mm-wave, and $\gamma$-ray bands, we constrain the $\gamma$-ray production region, probe the magnetic field topology, and investigate the spectral evolution in the innermost jet regions during this exceptional flare.\\
For our calculations, we adopt the following cosmological parameters: $\Omega_\mathrm{M}=0.27$, $\Omega_{\Lambda}=0.73$, $\rm H_{0}=71$\,km~s$^{-1}$~Mpc$^{-1}$ (similar to those used by \citealt[and references therein]{2016AJ....152...12L}), which result in a luminosity distance $D_{L}=5.489$\,Gpc and a linear-to-angular size conversion of 7.7\,pc/mas for the redshift of z=0.859.

\section{Observations and data reduction}
\label{sec:data}

\subsection{Data acquisition}

\begin{table*}[h!]
\centering
\small
\caption{VLBI observational and polarimetric parameters of \object{TXS~2013+370} on 11 February 2021}
\begin{tabular}{cccccccccccccccc}
\hline
\hline
Freq. & Array & $b_{\mathrm{maj}}$ & $b_{\mathrm{min}}$ & PA & rms & $S_{\mathrm{total}}$ & $\sigma_S$ & $P$ & $\sigma_P$ & $\sigma_{\mathrm{D\text{-}term}}$ & $m$ & $\Delta m$ & $\chi$ & $\Delta\chi$ \\
(GHz) &       & (mas) & (mas) & ($^\circ$) & (mJy/b.) & (Jy) & (Jy) & (mJy) & (mJy) & (mJy) & (\%) & (\%) & (deg.) & (deg.)\\
(1) & (2) & (3) & (4) & (5) & (6) & (7) & (8) & (9) & (10) & (11) & (12) & (13) & (14) & (15)\\
\hline
22 & VLBA$_{10}$+EB & 0.53 & 0.22 & $-12.5$ & 0.8 & 1.43 & 0.14 & 47 & 4.7 & 1.4 & 3.3 & 0.5 & 109 & 3.6\\
43 & VLBA$_{10}$+EB & 0.27 & 0.12 & $-13.7$ & 0.6 & 1.07 & 0.11 & 47 & 4.7 & 0.6 & 4.5 & 0.6 & 113 & 0.5\\
86 & VLBA$_7$ & 0.22 & 0.16 & $-3.6$ & 1.1 & 0.90 & 0.1 & 26 & 5.7 & 3.0 & 3.5 & 0.8 & 52 & 7.9\\
\hline
\end{tabular}
\vspace{5pt}
\tablefoot{Columns from left to right: (1) Observing frequency, (2) Participating antennas (note that at 86\,GHz, the VLBA stations at Hancock, Los Alamos, and St. Croix did not participate in the observations), (3)-(5) Beam parameters: major axis, minor axis, and position angle, (6) Image noise level, (7)-(8) Total intensity and its uncertainty, (9)-(10) Polarized flux and its uncertainty, (11) Systematic D-term calibration error, (12)-(13) Fractional polarization and its uncertainty, (14)-(15) Polarization angle and its uncertainty.}
\label{table:obs_par}
\end{table*}

VLBI observations of \object{TXS~2013+370} were performed on 11 February 2021 following elevated $\gamma$-ray emission from the source since 6 December 2020. Our VLBI data sets include polarimetric observations at 22 and 43\,GHz with the full VLBA array plus Effelsberg, and at 86\,GHz with seven VLBA antennas. The data were recorded for 8 hours in two polarizations (left and right circular polarized, LCP and RCP) with a 64\,MHz bandwidth per polarization, split into four intermediate frequency (IF) bands of 16 MHz each. The observations followed a standard VLBI calibration strategy, alternating between the target source, \object{TXS~2013+370} and the gain calibrators \object{3C\,345}, and \object{BL Lacertae}. The details of the aforementioned VLBI observations are summarized in Table~\ref{table:obs_par}.

\begin{figure}[h!]
\includegraphics[width=\columnwidth]{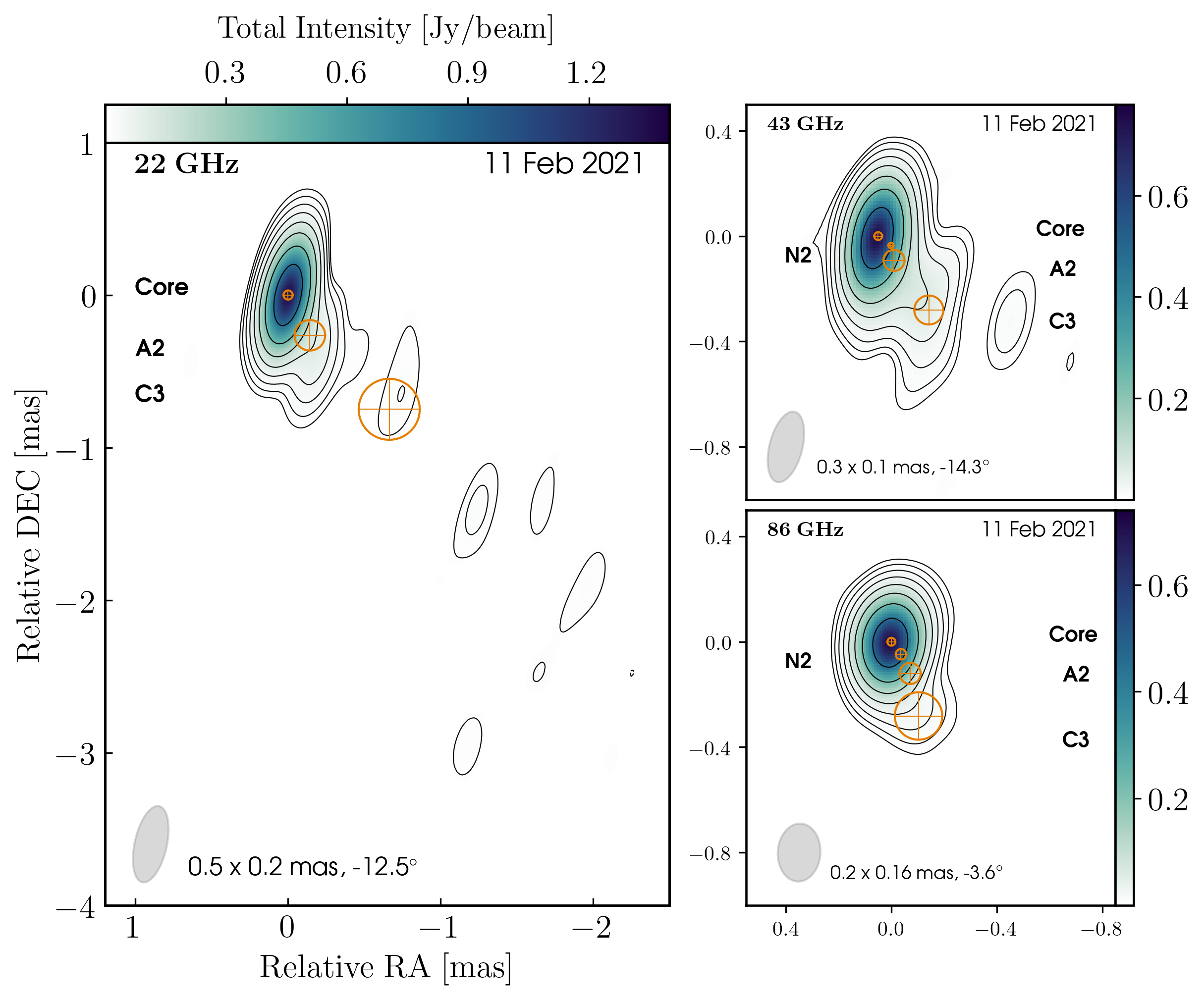}
\caption{Total intensity images of \object{TXS~2013+370} from 11 February 2021. Left: 22\,GHz; top-right: 43\,GHz; bottom-right: 86\,GHz. Contours are at 0.5, 1, 2, 4, 8, 16, 32, and 64\% of each panel’s peak (22: 1.33\,Jy/beam; 43: 0.75\,Jy/beam; 86: 0.71\,Jy/beam). Restoring beams are shown as gray ellipses (parameters in Table~\ref{table:obs_par}). Orange circles mark MODELFIT Gaussian centroids. The 43\,GHz map resolves a new knot (N2) near the core; at 86\,GHz the compact core and components A2, C3, and N2 are clearly detected.
}
\label{fig:imaging}
\end{figure}

\subsection{Calibration}

The data reduction was performed using the National Radio Astronomy Observatory's (NRAO) Astronomical Image Processing System (\texttt{AIPS}; \cite{1990apaa.conf..125G}). \texttt{AIPS} is a widely used software package for VLBI calibration and was used in the present work for all three frequencies: 22\,GHz, 43\,GHz, and 86\,GHz. The calibration procedure starts with the correction of phase errors in the cross-correlation data introduced by the sampler, followed by a parallactic angle correction of the phases. After the initial corrections, we performed manual phase calibration to determine the delay, phase errors, and rate corrections using fringe-fitting per baseline and high signal-to-noise ratio scans. After achieving phase alignment across the observing band and time for each baseline, global fringe fitting was performed to solve for the residual delays and phase errors with respect to the reference antenna. With the phases calibrated, we proceeded to visibility amplitude calibration. After updating the $T_{sys}$ and gain tables in \texttt{AIPS}, we performed absolute flux calibration, converting the observed visibility data into flux density values. We then applied amplitude calibration, incorporating corrections for atmospheric opacity, derived using the measured system temperatures and the gain-elevation curves of each telescope. The polarization alignment was performed using \texttt{RLDLY} task on a bright scan of the source.
The removal of instrumental polarization leakage from the data, also known as D-terms \citep{1994cers.conf..207L}, was performed via the task \texttt{LPCAL} in \texttt{AIPS}, following the process described in detail, along with the absolute EVPA calibration, in the Appendix \ref{ap:pol_cal}

\subsection{Imaging and model-fitting}

The total intensity and polarization images of \object{TXS~2013+370} at 22\,GHz, 43\,GHz, and 86\,GHz were produced using \texttt{Difmap} \citep{1997ASPC..125...77S}. The amplitude and phase calibrated data were imported into \texttt{Difmap}, where spurious data points were flagged and removed following careful inspection of the visibility amplitudes and phases. Images were created using the \texttt{CLEAN} deconvolution algorithm, an iterative procedure implemented in \texttt{Difmap}, producing three radio maps at the respective frequencies (see Fig.\ref{fig:imaging}).
For all three maps, the jet brightness distribution was parameterized using the \texttt{MODELFIT} algorithm in \texttt{Difmap}. This procedure fits two-dimensional Gaussian components to the fully calibrated visibility data, providing a quantitative description of the flux density distribution along the jet direction. Component uncertainties were formally computed based on the local signal-to-noise ratio (S/N) in the image surrounding each component \citep{1999ASPC..180..301F,2005astro.ph..3225L,2012A&A...537A..70S}. However, for the flux density uncertainty, we adhered to more conservative criteria, following the methodology described in \cite{2024A&A...682A.154T, 2025A&A...700A..16T}, and references therein, as the formal errors obtained via this method appeared too small. All parameters of the fitted Gaussian components are provided in Table \ref{table:components}.
Given that \object{TXS~2013+370} lies very close to the Galactic plane, interstellar scattering effects could potentially affect the radio images due to the high column density of the interstellar medium along the line of sight. However, as demonstrated by \citet{refId0} through detailed analysis of interstellar scattering effects on VLBI data, such effects are dominant only at longer cm-wavelengths for \object{TXS~2013+370}, with a threshold frequency of approximately 10\,GHz. Therefore, our millimeter-wavelength observations are essentially unaffected by interstellar scattering, ensuring reliable morphological analysis of the innermost jet regions.

\begin{table}[H]
\setlength{\tabcolsep}{4pt} 
\centering
\caption{Model-fitting parameters for \object{TXS~2013+370}.}
\begin{tabular}{@{\hskip 2pt}c@{\hskip 4pt}c@{\hskip 4pt}c@{\hskip 4pt}c@{\hskip 4pt}c@{\hskip 4pt}c@{\hskip 2pt}}
\hline
\hline
ID & Freq. & $S$ & $r$ & $\theta$ & FWHM \\
 & (GHz) & (mJy) & ($\mu$as) & ($^\circ$) & ($\mu$as) \\
(1) & (2) & (3) & (4) & (5) & (6)\\
\hline
& 86  & 210 $\pm$ 20 & -- & -- & 30 $\pm$ 3 \\
& 43  & 390 $\pm$ 40 & -- & -- & 30 $\pm$ 3 \\
\multirow{-3}{*}{\vspace{-1.0ex}Core} & 22 & 1320 $\pm$ 130 & -- & -- & 50 $\pm$ 5 \\
\hline
& 86  & 550 $\pm$ 60 & 50 $\pm$ 40 & -142.4 $\pm$ 0.6 & 40 $\pm$ 4 \\
\multirow{-2}{*}{\vspace{-0.5ex}N2} & 43 & 330 $\pm$ 30 & 40 $\pm$ 40 & -126.3 $\pm$ 0.6 & 20 $\pm$ 2 \\
\hline
& 86 & 10 $\pm$ 1 & 120 $\pm$ 40 & -149.4 $\pm$ 0.3 & 60 $\pm$ 6 \\
& 43 & 190 $\pm$ 20 & 110 $\pm$ 40 & -146.9 $\pm$ 0.3 & 80 $\pm$ 8 \\
\multirow{-3}{*}{\vspace{-1.0ex}A2} & 22 & 140 $\pm$ 10 & 350 $\pm$ 70 & -151.6 $\pm$ 0.2 & 200 $\pm$ 20 \\
\hline
& 86 & 40 $\pm$ 4 & 290 $\pm$ 40 & -159.9 $\pm$ 0.1 & 140 $\pm$ 10 \\
& 43 & 100 $\pm$ 10 & 340 $\pm$ 40 & -145.6 $\pm$ 0.1 & 110 $\pm$ 10 \\
\multirow{-3}{*}{\vspace{-1.0ex}C3} & 22 & 10 $\pm$ 1 & 1000 $\pm$ 80 & -138.5 $\pm$ 0.1 & 400 $\pm$ 40 \\
\hline
\end{tabular}
\vspace{5pt}
\tablefoot{Columns from left to right: (1) Component ID, (2) Observing frequency, (3) Flux density, (4) Radial distance from the core, (5) Position angle, (6) Component size (FWHM).}
\label{table:components}
\end{table}

\subsection{Single-dish radio light curves and \textit{Fermi}-LAT data}

\begin{figure}[!t]
\centering
\includegraphics[width=0.95\columnwidth]{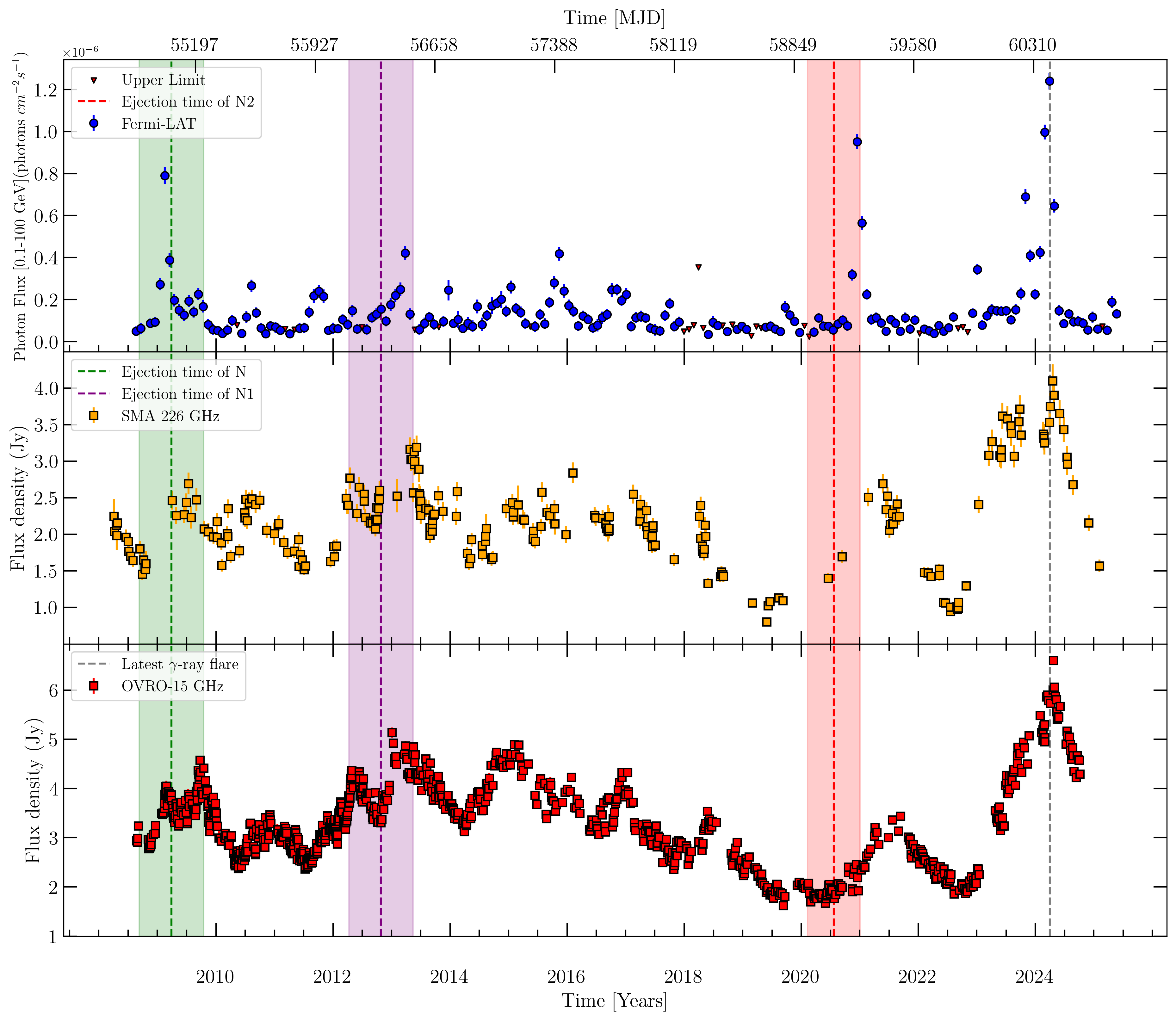}
\caption{Light curves of the blazar \object{TXS~2013+370} at different frequencies. From top to bottom: \textit{Fermi}-LAT 0.1-100\,GeV with 1-month binning, flux plotted vs. time for 226\,GHz SMA, and 15\,GHz OVRO. The vertical dashed lines indicate the estimated ejection times of the components N, N1 \citep{refId0}, and the new knot N2, respectively, whereas the width of the shadow areas designates the uncertainty of these estimations. For N2, the uncertainty is based on the uncertainty of component A1 as it was found in \cite{refId0}. The gray dashed line indicates a new flaring activity in the source.}
\label{fig:light_curves}
\end{figure}

Single-dish observations were provided by the 40-m telescope of the Owens Valley Radio Observatory (OVRO) at 15\,GHz, and the 8-element Submillimeter Array (SMA)\footnote{http://sma1.sma.hawaii.edu/callist/callist.html} at 226\,GHz. The $\gamma$-ray data were obtained by the \textit{Fermi}-LAT, the forth LAT source catalog
(4FGL; \cite{Abdollahi_2023}), with 1-month binning in the 0.1--100\,GeV energy range. All measurements span approximately the same period from 2008 to 2025. As shown in the multi-frequency light curves, presented in Fig.~\ref{fig:light_curves}, the source exhibited significant activity during our monitoring period, with two major flaring episodes occurring in 2021 and 2024. Our VLBI observations were conducted during the 2021 flare, which is clearly visible across all frequency bands. 

\section{Data analysis and results}
\label{sec:analysis_results}

\begin{figure}
\centering
\includegraphics[width=0.24\textwidth]{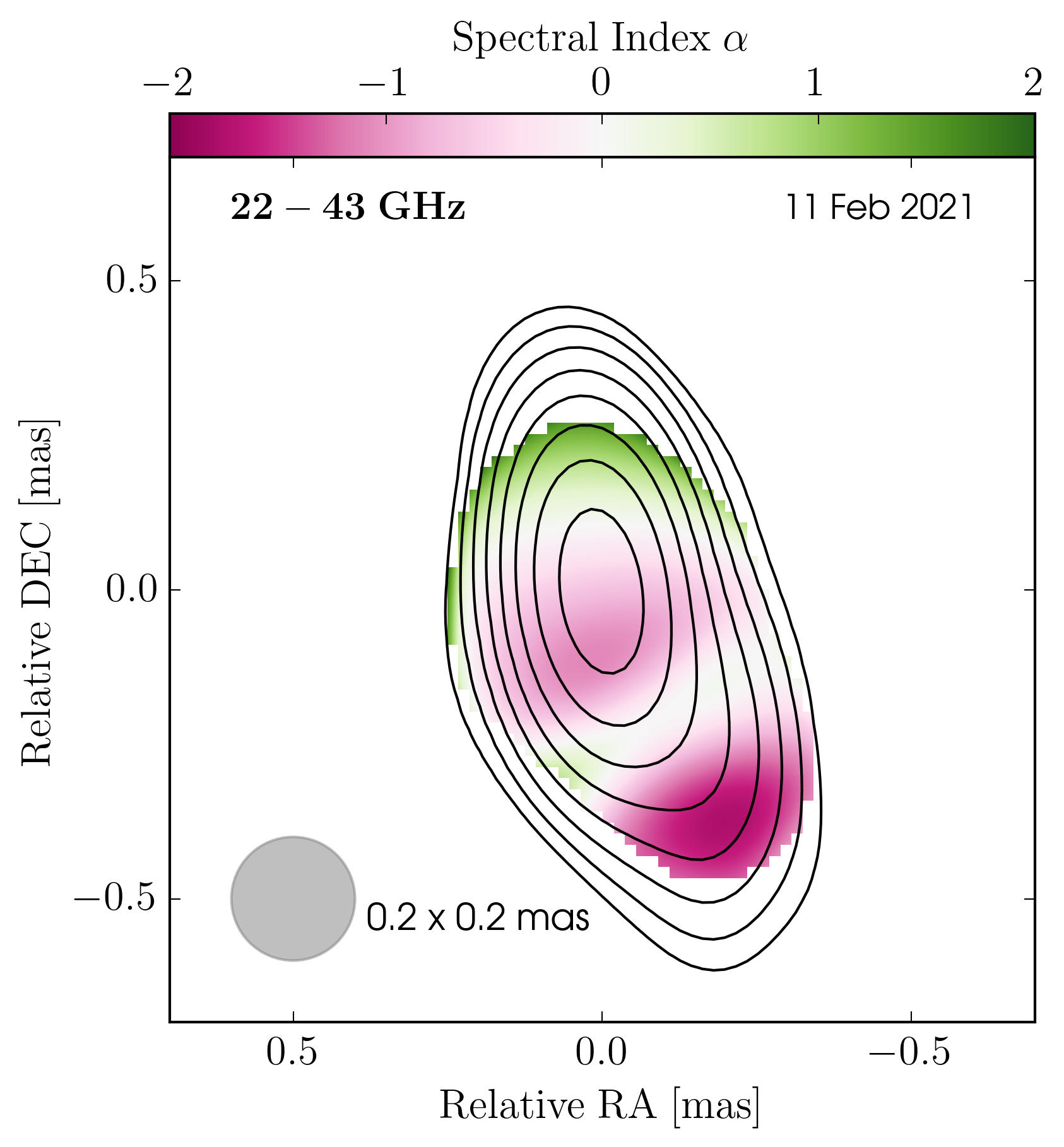}
\includegraphics[width=0.24\textwidth]{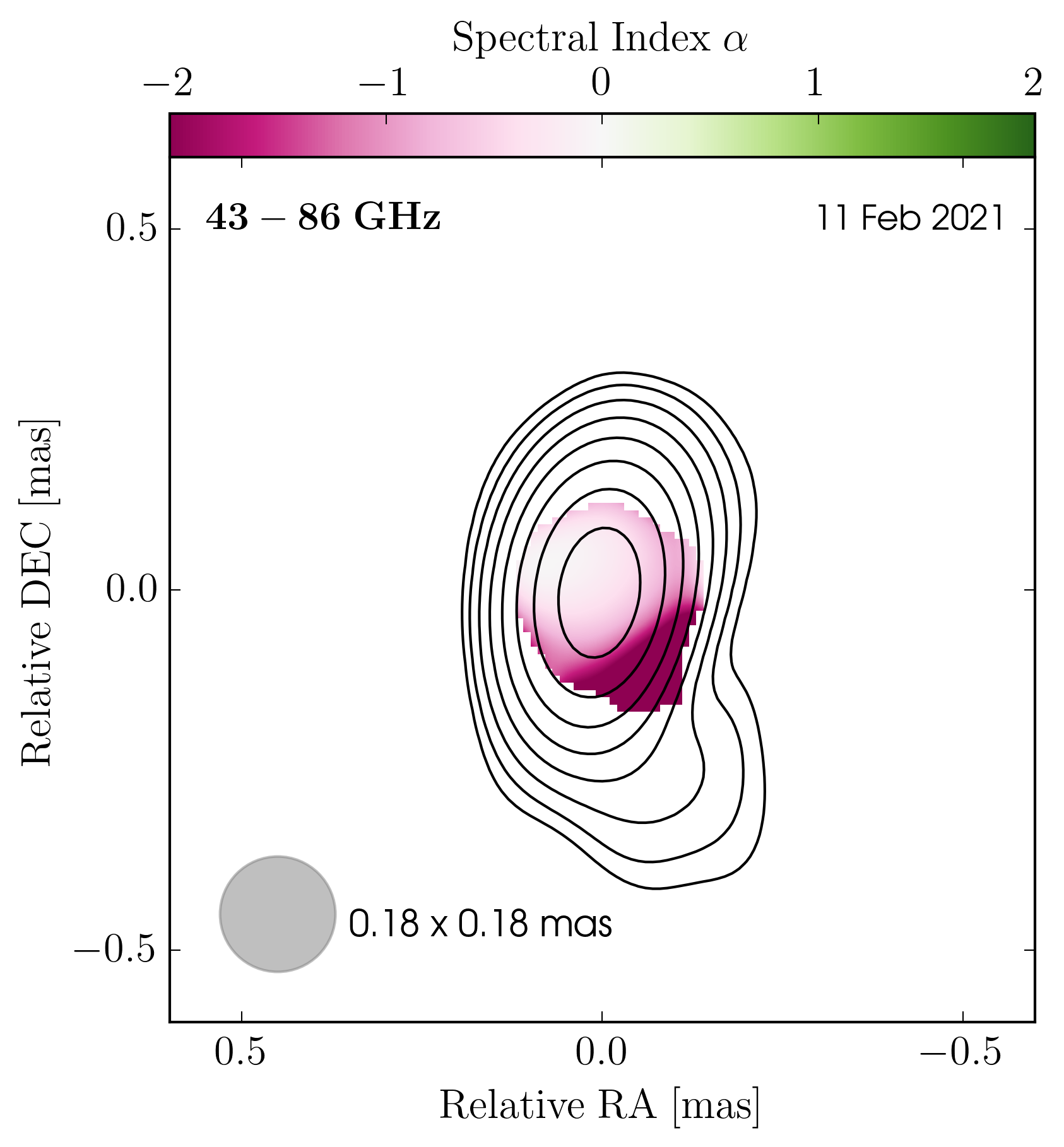}
\caption{Spectral index distributions of \object{TXS~2013+370}. The contour levels are set to 0.5, 1, 2, 4, 8, 16, 32, and 64\% Jy/beam of the peak flux density of the highest frequency map in the pair (see Table: \ref{table:obs_par}) and represent the total intensity contours. All the images are convolved with a common beam that was set equal to the equivalent circular beam $b=(b_{max}b_{min})^{1/2}$ of the highest frequency, and each frequency pair was aligned using a 2D cross-correlation analysis. Left: 22-43\,GHz frequency pair.  Right: 43-86\,GHz frequency pair.}
\label{fig:Spectral}
\end{figure}

\subsection{Source structure and jet components}

Our multi-frequency VLBI observations at 22, 43, and 86 GHz reveal \object{TXS~2013+370} as a compact, core-dominated source with a curved jet structure extending southwestward from the bright core region. The total intensity images (Figure \ref{fig:imaging}) show a well-defined morphology consisting of a dominant core and several distinct jet components, with the overall jet structure becoming increasingly well-resolved at higher frequencies. In addition to the N2, we identify stationary features A2 and C3, whose positions and flux densities are consistent with the quasi-stationary components reported by \cite{refId0}. At 22\,GHz, the source displays a relatively simple structure with a bright, unresolved core and extended emission along the jet direction. The core dominates the total flux density with 1.32\,Jy (92\% of the total flux), while the jet components contribute the remaining emission. The 43\,GHz observations reveal additional structural details. Most significantly, these observations enable the detection of component N2, a newly emerged feature located approximately 60\,$\mu$as from the VLBI core. This component exhibits a flux density of 330 $\pm$ 30\,mJy at 43\,GHz and is likely associated with the enhanced multi-wavelength activity observed during our monitoring period. Such associations between newly emerged VLBI components and $\gamma$-ray flares are common in blazars, with $\sim$83\% of GeV outbursts coinciding with new superluminal features in the VLBA-BU-BLAZAR monitoring program \citep{2017ApJ...846...98J}. At the highest frequency (86\,GHz), component N2 is clearly visible with a flux density of 550 $\pm$ 60\,mJy, while the core becomes more compact at this frequency, containing 210 $\pm$ 20\,mJy. 

The Gaussian model-fitting analysis (presented in Table \ref{table:components}) helps us to quantify the jet curved morphology with position angles ranging from approximately -126$^{\circ}$ for component N2 to -160$^{\circ}$ for the more distant component C3. Component sizes increase systematically with distance from the core (20-400\,$\mu$as), consistent with adiabatic jet expansion. The flux density distribution among components varies significantly with frequency, reflecting different opacity conditions across the jet. At 86\,GHz, where synchrotron opacity effects are minimal, component N2 contributes a significant fraction of the total emission after the core, taking the leading role in the flaring episode.

\subsection{Spectral decomposition}

We reconstructed the spectral index along the jet using quasi-simultaneous image pairs at 22–43, and 43–86\,GHz, adopting the definition $S_\nu \propto \nu^{+\alpha}$. For each frequency pair, all maps are convolved by a common restoring beam, corresponding to the equivalent circular beam of the higher frequency, following standard practice (e.g., \cite{2024A&A...682A.154T}), and a pixel size of $18\,\mu$as. Image alignment was performed via 2D cross-correlation \citep{Gabuzda2004}, in which the lower-frequency map was held fixed while the higher-frequency map was systematically shifted pixel-by-pixel to maximize the cross-correlation 
coefficient between the two intensity distributions. This approach is appropriate when optically thin jet features suitable for direct component-based alignment are not available or when core shifts are below the resolution limit. For TXS~2013+370, \citet{refId0} demonstrated that core shifts between 15 and 86\,GHz are $\lesssim$0.1\,mas, well below our beam sizes, justifying a correlation-based alignment method. The resulting shifts are $(\Delta x,\Delta y)=(0,+1)$ pixels at 22–43\,GHz and $(-2,+1)$ pixels at 43–86\,GHz. Visual inspection confirmed physically consistent spectral structures in the aligned maps. The resulting maps are shown in Fig.~\ref{fig:Spectral}.

The spectral index maps show relatively uniform flat spectra in the core region across all frequency pairs, with some steeper indices appearing in the outer jet regions where signal-to-noise ratios are lower. To quantify the core spectral properties, we computed a representative value by applying a mask (64\% of the peak of the highest frequency map in the pair) to calculate the average of the indices within the first contour, with uncertainties following \citet{bartolini2025}. From the results, we notice that the VLBI core exhibits spectral indices of $\alpha_{\rm core}=-0.43\pm0.01$ at 22–43\,GHz and $-0.51\pm0.11$ at 43–86\,GHz, with a mean value $\alpha \gtrsim -0.5$, indicating a flat spectrum characteristic of partially self-absorbed emission. The consistency of these values across both frequency pairs suggests stable spectral properties during the flare epoch.

\subsection{Polarization and Faraday rotation}
\label{sec:pol}

\begin{figure*}
\centering
\includegraphics[width=0.32\textwidth]{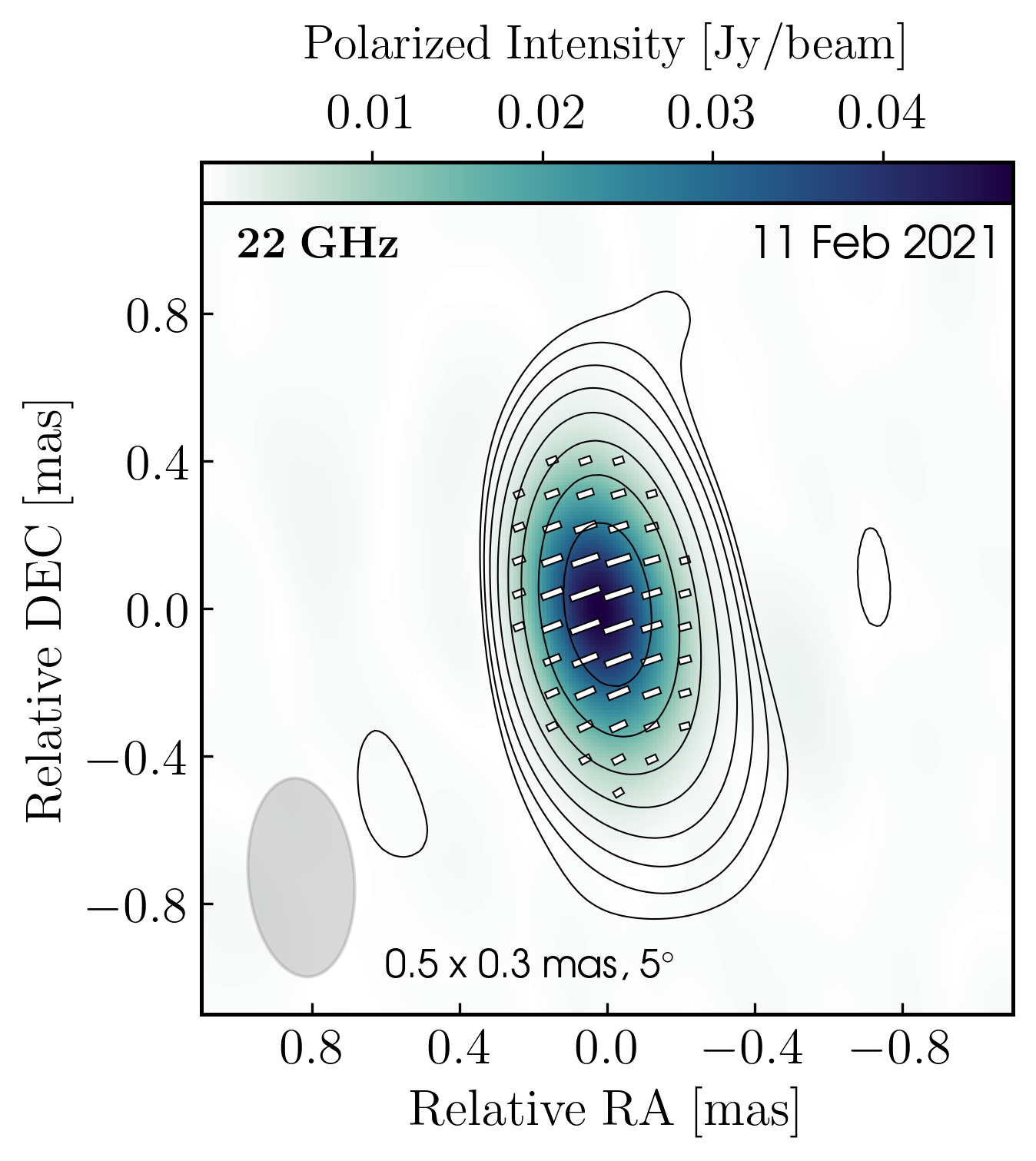}\hfill
\includegraphics[width=0.33\textwidth]{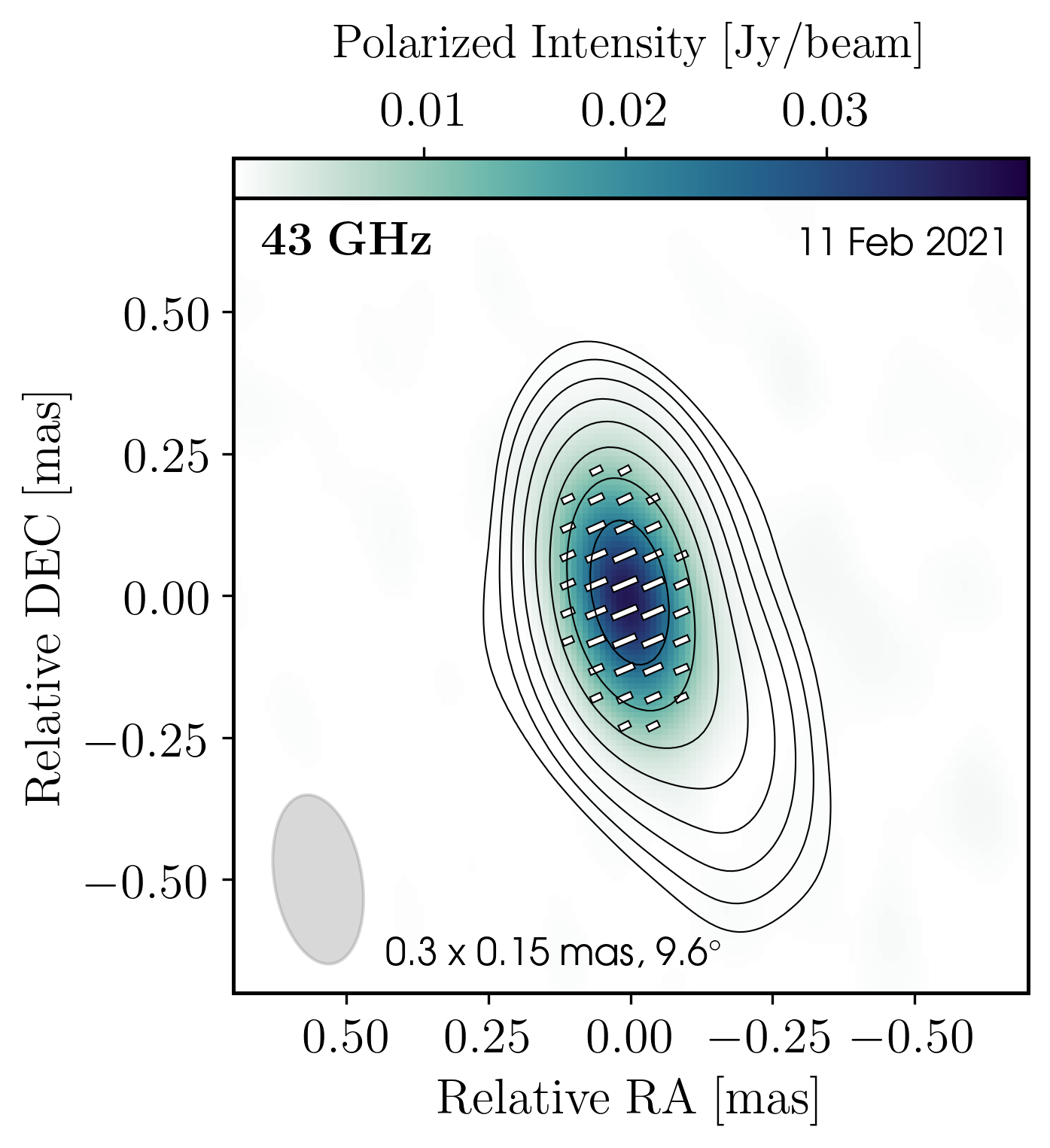}\hfill
\includegraphics[width=0.32\textwidth]{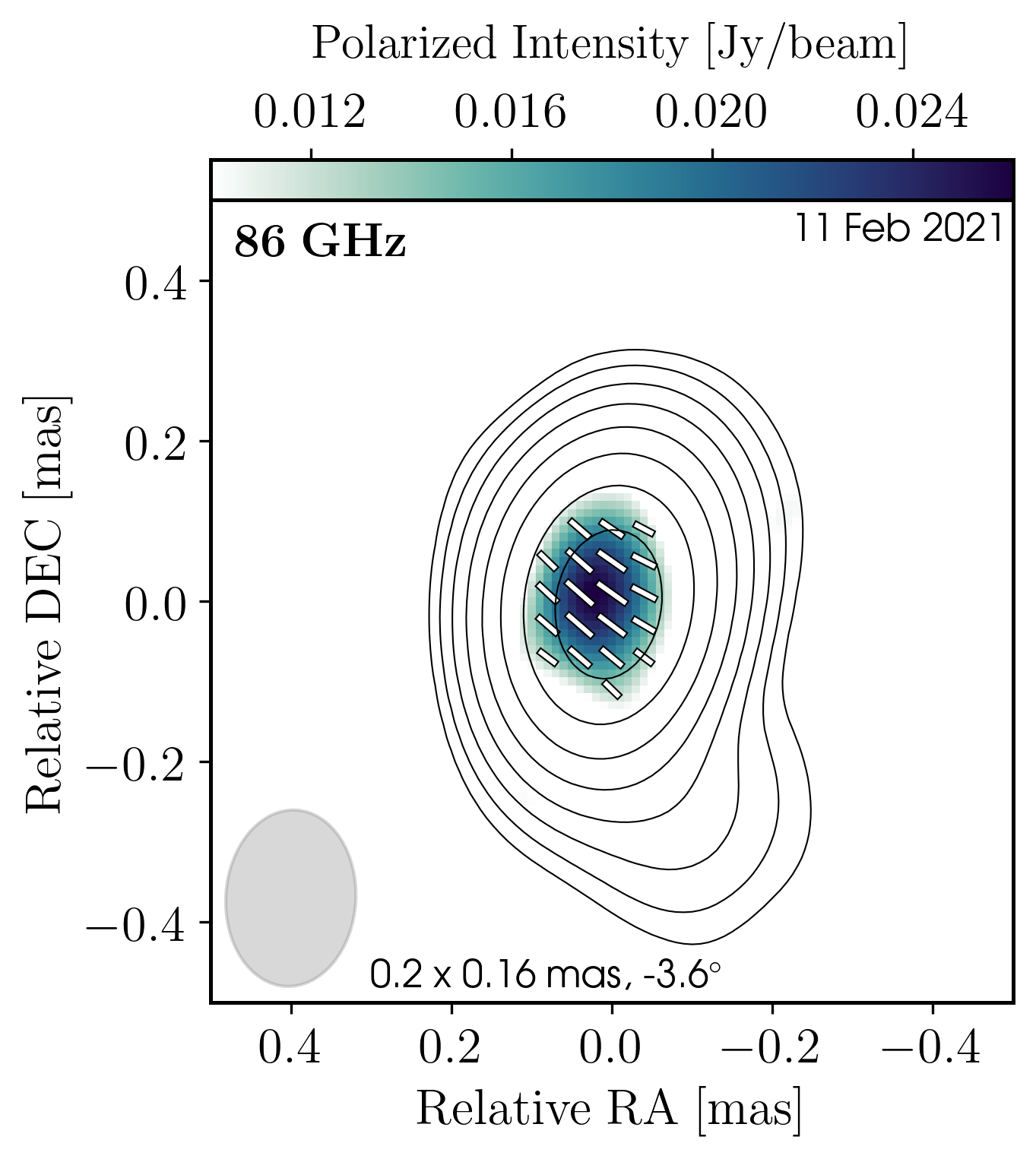}
\caption{Polarization images of \object{TXS~2013+370} observed on 11 February 2021. Left: 22\,GHz polarized-intensity map with overplotted EVPA vectors (white sticks). Middle: 43\,GHz polarization image. Right: 86\,GHz polarization image. Contours are at 0.5, 1, 2, 4, 8, 16, 32, and 64\% of each panel’s peak (22\,GHz: 1.31\,Jy\,/beam; 43\,GHz: 0.80\,Jy\,/beam; 86\,GHz: 0.70\,Jy\,/beam) and represent the total intensity contours. The restoring beam is shown as a gray ellipse in the lower-left corner.
}
\label{fig:polarization}
\end{figure*}

Polarimetric VLBI data of the source at 22\,GHz, and for the first time at 43 and 86\,GHz (calibration in Appendix~\ref{appendix:Dterms}), reveal a core with strong, coherent linear polarization and a well-aligned magnetic field (Fig.~\ref{fig:polarization}). The fractional linear polarization reaches $3.3\pm0.5\%$ (22\,GHz), $4.5\pm0.6\%$ (43\,GHz), and $3.5\pm0.8\%$ (86\,GHz), with EVPAs of $109\pm4^{\circ}$, $113.0\pm0.5^{\circ}$, and $52\pm8^{\circ}$, respectively (Table~\ref{table:obs_par}). EVPA uncertainties follow \citet{Hovatta_2012}, calculated by combining in quadrature the single-dish reference angle error ($\sim3.6^\circ$ for 22 GHz, $\sim0.5^\circ$ for 43 GHz, and $\sim7.9^\circ$ for 86 GHz) and the propagated error, with the latter derived from the rms of the $Q$/$U$ maps and a 10\% calibration term.

A notable feature in our images is a sudden $\sim50^{\circ}$ EVPA rotation at 86\,GHz, which changes the inferred magnetic field orientation from roughly parallel to nearly perpendicular to the jet axis. This behavior may indicate a change in field geometry (e.g., helical with toroidal-to-poloidal dominance; \citealt{Gabuzda2004,ONeill2019}), though opacity effects cannot be excluded. To test this, we corrected for Faraday rotation by calculating the rotation measure (RM) for each pixel \citep[e.g.,][]{2013MNRAS.429.3551A}, using:

\begin{equation}
\chi(\lambda)=\chi_0+\mathrm{RM}\, \cdot \lambda^2 
\label{eq:rm}
\end{equation}

where $\chi(\lambda)$ is the observed EVPA, $\chi_0$ is the intrinsic EVPA, and $\lambda$ is the observing wavelength. We derived the pixel-based RM by linearly fitting $\chi$ versus $\lambda^2$ for each pixel, with explicitly unwrapping $n\pi$ ambiguities following \cite{Hovatta_2012}. Only pixels with polarized intensities exceeding $3\sigma_P$ were used, and a relaxed $\chi^2$ threshold ensured robust fits across the core. In addition, an integrated RM value for the core was obtained by fitting the frequency-averaged U and Q, yielding a result consistent with the pixel-based analysis.

\begin{figure*}
\centering
\includegraphics[width=0.29\textwidth]{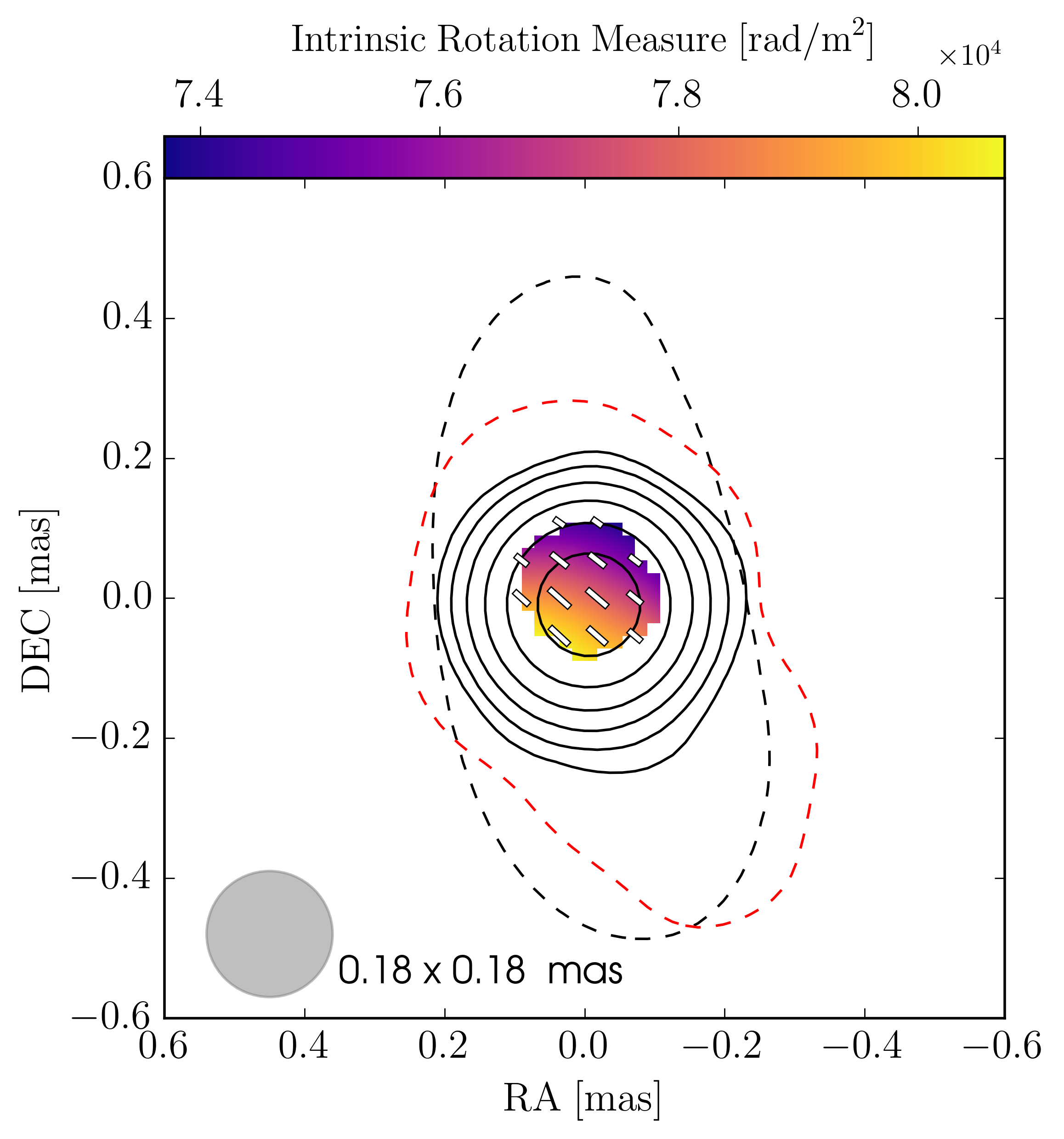}
\includegraphics[scale=0.46]{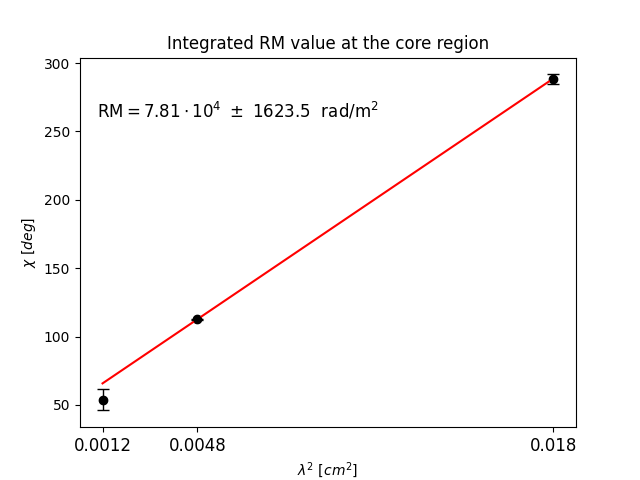}
\includegraphics[width=0.29\textwidth]{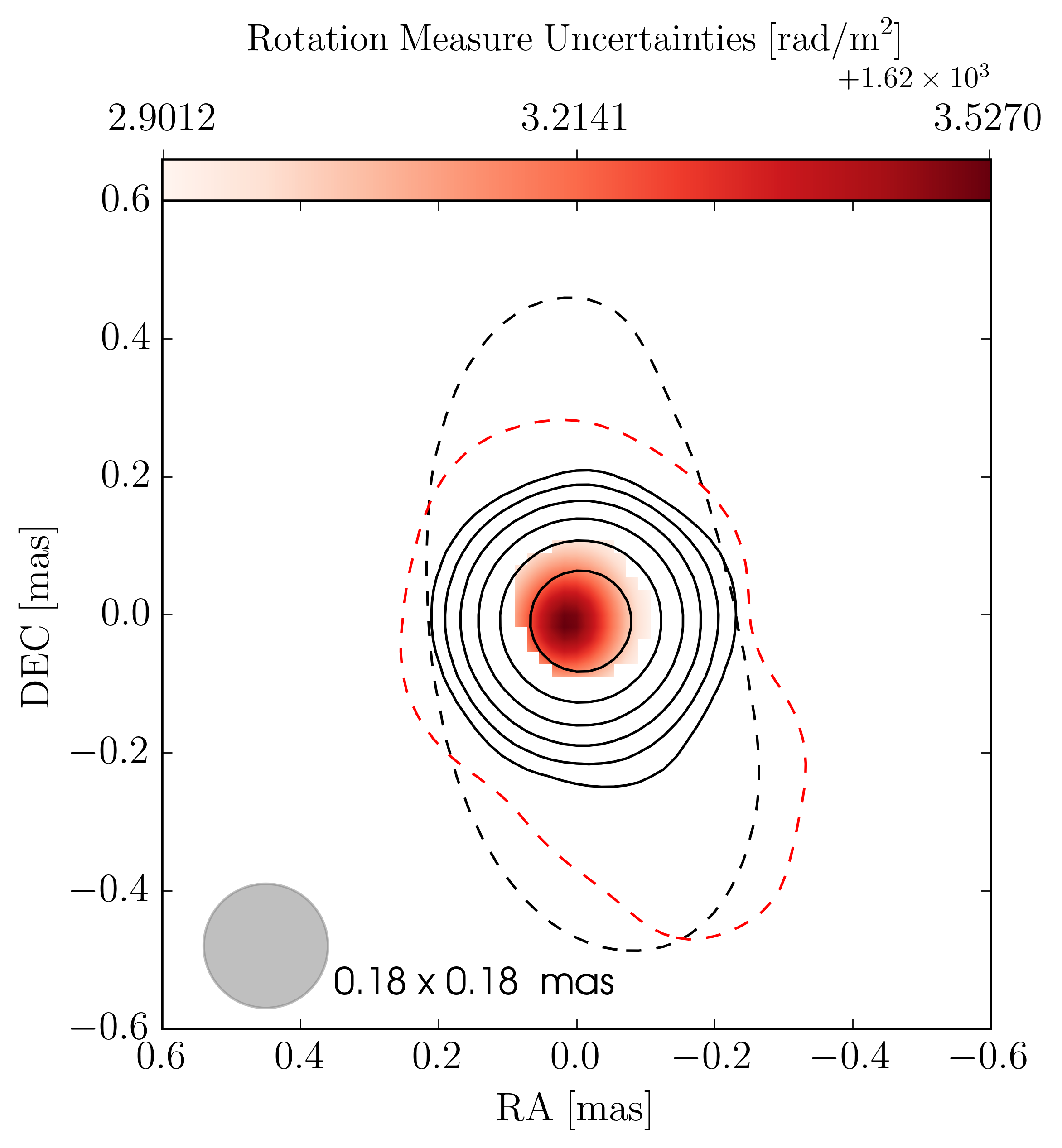}
\caption{Faraday rotation results for the blazar \object{TXS~2013+370}. 
Left: Dashed lines correspond to the outermost total intensity contours of the 22\,GHz (black) and 43\,GHz (red) maps. Solid contours are set to 2, 4, 8, 16, 32, and 64\% of the 86\,GHz peak polarized intensity (0.70\,Jy\,/beam). All maps are convolved with a common beam, the equivalent circular beam of the 86\,GHz map, shown as a gray circle in the bottom-left corner. 
Middle: Linear fit of Eq.~\ref{eq:rm} after unwrapping the $n\pi$ ambiguities, showing EVPA rotation from 86\,GHz to 22\,GHz; the slope yields an integrated RM of $(7.81\pm0.16)\times10^{4}$\,rad\,m$^{-2}$ for the core region. 
Right: Pixel-based map of RM uncertainties from the covariance matrix of the per-pixel least-squares fits.}
\label{fig:Faraday}
\end{figure*}

The RM map presented in Fig.~\ref{fig:Faraday} reveals a high and uniform rotation measure of $\mathrm{RM}=(7.81\pm0.16)\times10^{4}\ \mathrm{rad\,m^{-2}}$, with uncertainties from the covariance matrix of the per-pixel fits lying within a narrow range across the core region. These values significantly exceed typical blazar-core RMs of $10^2$–$10^4\ \mathrm{rad\,m^{-2}}$ \citep[e.g.,][]{2012AJ....144..105H,2013MNRAS.429.3551A}. The RM distribution shows high uniformity with no strong transverse gradients, and is accompanied by a $\sim50^{\circ}$ EVPA rotation between 22-43\,GHz and 86\,GHz. The physical interpretation of these findings is discussed in Section~\ref{subsec:faraday_interpretation}.

\subsection{Multi-band variability }
\label{subs:multi_band_var}

To investigate the connection between gamma-ray and radio variability, we performed a comprehensive correlation analysis. We first established the reality of gamma-radio associations across the complete 16-year observational record (2008-2025, Figure~\ref{fig:light_curves}). We defined gamma-ray flares as periods where the \textit{Fermi}-LAT flux exceeds $3 \times 10^{-7}$\,photons\,cm$^{-2}$\,s$^{-1}$ (approximately 3$\times$ the quiescent level) for $> 30$\,days, identifying 5 prominent events. Radio flares were defined as enhancements in the OVRO 15\,GHz light curve exceeding 3.0\,Jy (approximately 1.3$\times$ the baseline level of $\sim 2.3$\,Jy) sustained for $> 15$\,days, yielding 10 distinct radio flares. We used OVRO data for this analysis as the SMA 226\,GHz observations exhibit significantly higher variability noise and sparser observational cadence, which obscures clear flare identification. 

To assess statistical significance, we performed Monte Carlo simulations by 
randomly shuffling the 5 gamma-ray flare occurrence times (keeping 10 radio flare times fixed) across the 16-year observation window. For each of 10,000 trials, we counted how many gamma-ray flares had an associated radio flare within $\pm$200-day windows. Only 130 trials showed $\geq$4 associations (matching our observed 4/5 success rate), yielding $p = 0.013$. This demonstrates that the observed gamma-radio correlation is unlikely to arise from random temporal overlap. Gamma-ray flares were defined as periods with flux $> 3 \times 10^{-7}$\,photons\,cm$^{-2}$\,s$^{-1}$ sustained for $>30$\,days, radio flares as flux $>3.0$\,Jy sustained for $>15$\,days. These thresholds were chosen to capture major outbursts clearly visible in Figure~\ref{fig:light_curves} while avoiding identification of variability noise as flares.

We then applied quantitative cross-correlation analysis using the discrete cross-correlation function (DCCF; \citealt{1988ApJ...333..646E}) between the \textit{Fermi}-LAT and OVRO\,15\,GHz light curves on a 6\,d lag grid (requiring $\geq 8$ pairs per bin) for the 2019-2022 interval, which provides homogeneous sampling and a single, coherent activity state. The amplitude significance was assessed with power spectral density (PSD) + probability density function (PDF), matched surrogate $\gamma$-ray light curves following \citet{2013MNRAS.433..907E}: we generated 5000 realizations, cross-correlated each with the 15\,GHz data, and derived the 95\% and 99\% confidence envelopes as a function of lag. 

\begin{figure}
\includegraphics[width=0.9\linewidth]{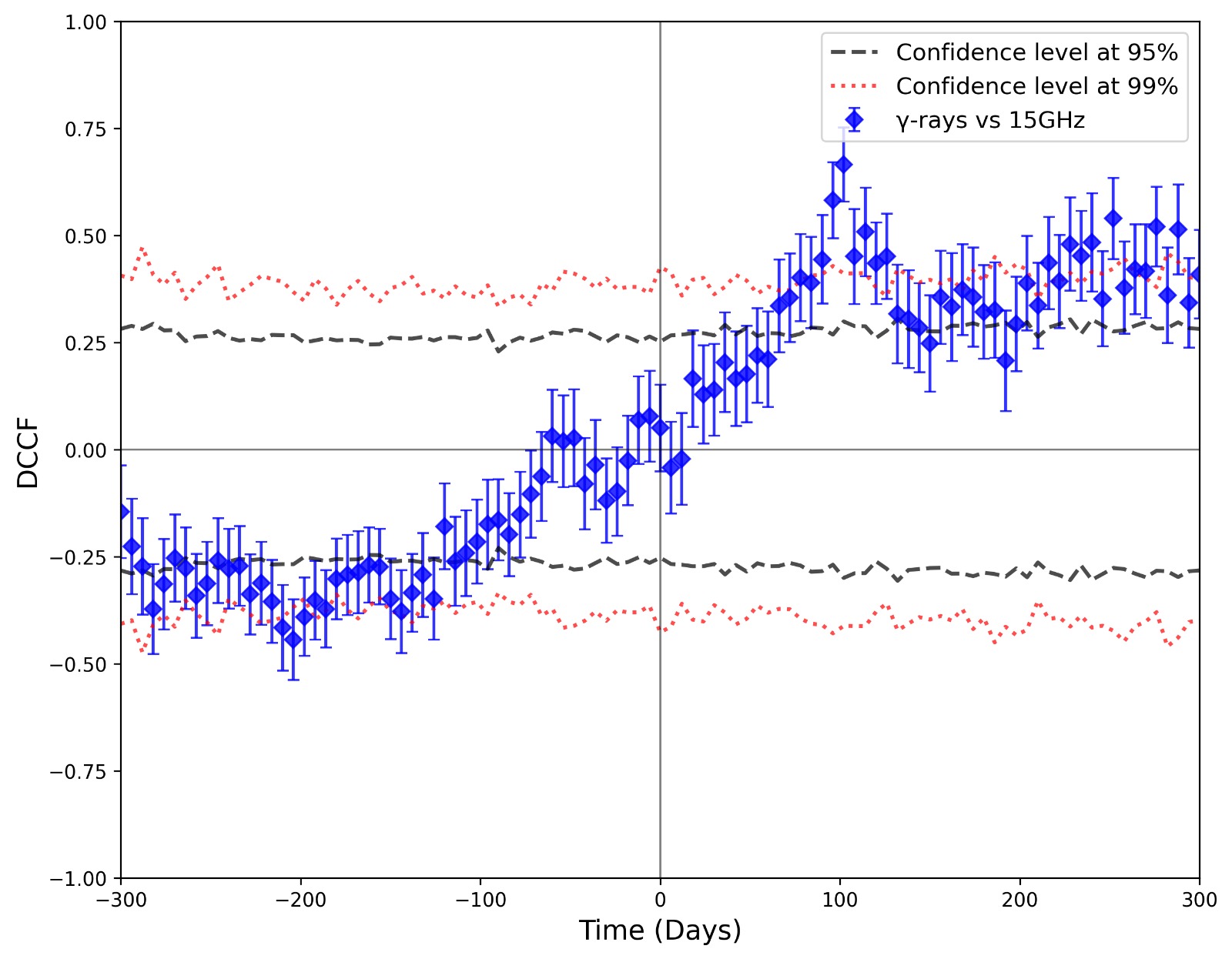}
\caption{DCF results between the $\gamma$-ray and 15\,GHz light curves.
Positive time lags indicate that $\gamma$-ray activity leads the activity in radio. The significance of the correlations is displayed by a dashed line for the 2$\sigma$ and by a dotted line for the $3\sigma$ level.}
\label{fig:gamma-15}
\end{figure}

The $\gamma$-15\,GHz DCCF shows a prominent peak of $r = 0.67 \pm 0.09$ 
at a lag of 102\,days, corresponding to a $7.7\sigma$ significance, well above the 99\% confidence envelope (derived from 1000 autoregressive AR(1) surrogate pairs), confirming a highly significant correlation. The time lag and its uncertainty were obtained via FR/RSS Monte Carlo resampling (e.g. \citealt{1998PASP..110..660P}): for each of 5000 realizations we perturbed the fluxes by their errors and bootstrap-resampled the time series, recomputed the DCCF, and measured a weighted centroid of the peak using points with $r \ge 0.9\,r_{\max}$ within a region of interest centered on the observed peak (half-width set by the DCCF FWHM, $\simeq 65$\,d, with a minimum of 60\,d). We adopt the median lag and the 16th-84th percentiles as the $1\sigma$ statistical interval and add in quadrature a small resolution floor that reflects the binning/cadence. This yields a $\gamma$-ray lead of $\Delta t = (102 \pm 12)$\,days (Fig.~\ref{fig:gamma-15}).

\section{Discussion}
\label{sec:discussion}

\subsection{Location of the $\gamma$-ray emission}
\label{subs:gamma_loc}

A strong correlation was found between the $\gamma$-ray and 15\,GHz variability, indicating that the high-energy activity leads the radio by $\Delta t = (102 \pm 12)$ days. Following \cite{Pushkarev_2010}, this observed time lag can be translated into the de-projected linear distance $\Delta r$ between the $\gamma$-ray and 15\,GHz emission zones, as:

\begingroup
\setlength{\abovedisplayskip}{1pt}\setlength{\belowdisplayskip}{6pt}
\begin{equation}\label{eq:distance}
\Delta r=\frac{\beta_{\rm app}c\,\Delta t}{\sin\theta\,(1+z)}\,,
\end{equation}
\endgroup
where $\beta_{\rm app}$ is the plasma apparent speed (in units of $c$), and $\theta$ is the jet viewing angle. Because our single-epoch data set does not allow a direct determination of $\beta_{\rm app}$ and $\theta$, we adopt the values from \cite{refId0} for the innermost moving component A1, $\beta_{\rm app}=4.2 \pm 0.5$ and $\theta=4.1^{\circ}\!\pm 0.2^{\circ}$. We emphasize that this speed represents the inner jet ($<$0.5\,mas from the core). While 15\,GHz VLBI monitoring shows higher speeds ($\beta_{\rm app} \sim 14$) at larger distances, \citet{refId0} demonstrated clear velocity stratification with $\Gamma \sim 6$ in the inner jet increasing to $\Gamma > 14$ at parsec scales, reflecting ongoing jet acceleration. Using outer-jet speeds would overestimate the $\gamma$-ray location ($\Delta r \sim 8$\,pc, placing emission at $\sim$6\,pc), inconsistent with particle energization in the acceleration zone. With these parameters and the measured lag, we obtain: $\Delta r_{\gamma-15} \;=\; (2.71 \pm 0.47)\ {\rm pc}\,$, where the uncertainty is calculated by propagating errors in quadrature from the lag measurement (12 d) and the uncertainties on $\beta_{\rm app}$ and $\theta$. Considering the reported jet geometry and the presence of a recollimation shock just downstream of the VLBI core \citep{refId0}, we adopt $R \le (2.05 \pm 0.97)$\,pc as the de-projected distance of the 15\,GHz core from the jet apex. The location of the $\gamma$-ray emission is then: $r_{\gamma} \equiv R - \Delta r = 2.05 - 2.71 = -0.66 \pm 1.08$\,pc. This formally negative value, while carrying large uncertainties that span from upstream of the apex to $\sim$0.42\,pc downstream, indicates that the $\gamma$-ray site likely lies between the jet apex and the 15\,GHz core, consistent with the observed $\gamma$-ray lead in the light curves. This is compatible with the $r_\gamma \sim 0.75 \pm 1.26$\,pc (or $\sim$1\,pc) location found by \citet{refId0} for the 2009 flare. On these scales, and given the FSRQ nature of \object{TXS~2013+370}, the relevant external seed photons for inverse Compton scattering are those of the BLR and the dusty torus. Therefore, we interpret these constraints as an indication that the $\gamma$-ray emission region likely lies within the BLR or exceeds it, entering the torus environment and reaching distances of $\sim0.42$ pc in the $1\sigma$ range. Our conclusions are not sensitive to the exact location; rather, they permit two plausible interpretations: 1) With a BLR size of $\sim 0.07$\,pc \citep{refId0}, the emission region is fully compatible with locations closer to the base of the jet, where optical and UV BLR-photons get up-scattered to high-energies via EC, and 2) Our inferred $\gamma$-ray site lies beyond or at the BLR edge, making the dusty torus the primary photon reservoir, with infrared photons intercepted by the jet and up-scattered to $\gamma$-rays also via EC. It is also highly possible that both the BLR and the dusty torus actively contribute to shaping the high-energy output of TXS 2013+370. Nonetheless, in \cite{refId0}, they constrained the location of the $\gamma$-ray emission on scales of 1–2 parsecs from the central engine, concluding that the dusty torus is the best candidate for providing a rich seed photon field. Additionally, in \cite{Kara_2012}, they showed that the SED modeling of TXS 2013+370 required the existence of an external radiation field with low temperatures ($T\sim10^2$ K) in order to find the best possible fit for the dominant emitting region, indicating an origin from cold dust. Taking into account the previous studies, it seems that the scale tips toward one scenario, favoring the dusty torus as the most possible seed photon reservoir. Our results are positioned in the intersection of these analyses, yielding strong insights that, indeed, the dusty torus is a possible seed photon field, but the BLR cannot be neglected, considering it an equal possibility.

\subsection{Temporal comparison of $\gamma$-ray emission sites and knots}
\label{subsec:2010v2021_compact}

Our two well-studied flares (in 2009 and late 2021) place the $\gamma$-ray dissipation region in the same inner-parsec neighborhood despite differing $\gamma$–15\,GHz lags. For the 2009 flare \citep{refId0}, the lag analysis located the $\gamma$-ray zone at $\sim$1\,pc downstream of the jet apex—beyond the BLR and naturally compatible with EC on dusty-torus photons \citep[e.g.,][]{2000ApJ...545..107B,2008MNRAS.386L..28G,2009ApJ...692...32D}. For the 2021 outburst, our DCCF yields a robust $\gamma$–15\,GHz lead of $(102\pm12)$\,d, implying a separation of $(2.71\pm0.47)$\,pc between the $\gamma$ site and the 15\,GHz core; combined with the upper limit to the 15\,GHz core–apex distance ($R_{15}\!\lesssim\!2.05\pm0.97$\,pc), this bounds the $\gamma$-ray location to $\sim$0–0.42\,pc (1$\sigma$) downstream of the apex. Thus, both epochs point to sub-parsec/parsec scales within the torus-dominated EC regime, with the current study also supporting the BLR domain as an equal possibility.

The differing lags we see between the two events are most naturally attributed to changes in the inner-jet state rather than relocations of the dissipation site. In 2021, we detected a compact near-core knot, a flat core-dominated spectrum, and coherent mm polarization signatures of enhanced particle and magnetic energy densities around the core. These conditions increase synchrotron opacity and shift the 15\,GHz core downstream, lengthening the $\gamma$–15\,GHz delay even if $r_\gamma$ is unchanged \citep[e.g.,][]{2012A&A...537A..70S}. By contrast, the shorter 2009 lag is consistent with a less opaque core and/or a flare that couples more directly to a downstream standing feature within the same spatial zone \citep[e.g.,][]{2016A&A586A60,2016JPhCS.718e2032R}. For context, simple peak-to-peak offsets during the 2024 activity (MJD$_\gamma\!=60398$, MJD$_{1\,\mathrm{mm}}\!=60418$, MJD$_{15\,\mathrm{GHz}}\!=60421$) give $\Delta t_{\gamma-1\,\mathrm{mm}}=20$\,d and $\Delta t_{\gamma-15\,\mathrm{GHz}}=23$\,d, indicative separations of $\simeq0.53$\,pc and $\simeq0.61$\,pc for the same geometry. Although sensitive to cadence and flare asymmetry, the frequency ordering and scales reinforce a stable inner-parsec $\gamma$-ray site and a recollimation-shock nature of the VLBI core.

Parsec-scale $\gamma$-ray zones similar to our result are widely reported: in the BL~Lacs \object{AO~0235+164} and \object{OJ~287}, major flares arise $\gtrsim$12–14\,pc downstream, co-spatial with the mm core/standing features \citep{Agudo2011b,Agudo2011a}; among FSRQs, \object{3C~454.3} links GeV outbursts to superluminal knots crossing the 43\,GHz core \citep{Jorstad2013}, \object{PKS~1222+216} shows VHE emission beyond the BLR \citep{Tavecchio2011}, and \object{PKS~1510$-$089} often favors pc-scale sites near the mm core, with episodes from multiple zones \citep{Orienti2013,Brown2013,Abdalla2021}; even for \object{3C~273}, rapid flares constrain the GeV region to $\lesssim$1.6\,pc \citep{Rani2013}. Collectively, these studies place many strong $\gamma$-ray outbursts $\sim$0.1–few\,pc from the apex, consistent with our findings for \object{TXS~2013+370}. Finally, decade-long multiwavelength/polarimetric monitoring of \object{CTA~102} indicates zero-lag correlations between $\gamma$-ray and optical emission over multiple cycles, implying co-spatiality between the optical emission zone and the seed photon reservoir for $\gamma$-ray production on year timescales \citep{Ma2025CTA102}.

\subsection{Core spectral properties and particle acceleration}
\label{subsec:core_spectrum}

The flat core spectral indices ($\alpha \gtrsim -0.5$) measured across both 22-43 and 43-86\,GHz frequency pairs indicate partially self-absorbed synchrotron emission in the VLBI core region during the 2021 flare. Such flat spectra are characteristic of inhomogeneous jets and may be shaped by a standing recollimation shock near the core or by the superposition of traveling shocks downstream \citep[e.g.,][]{1977AJ.....82..781M,Marscher2008}. The temporal association between the flat spectrum and the emergence of component N2 suggests ongoing particle injection and efficient acceleration 
during the flare.
The measured spectral indices are consistent with shock-accelerated electron distributions with power-law index $p\simeq2$2.2, corresponding to optically thin spectral indices $\alpha_{\rm thin}\simeq-0.5$ to $-0.6$ \citep{1978MNRAS.182..147B,1978ApJ...221L..29B}. While these values favor shock acceleration as the primary particle energization mechanism \citep{Marscher2008}, contributions from magnetic reconnection cannot be excluded \citep{Mimica2009,Sironi_2014}, particularly given the complex magnetic field structure revealed by our polarimetric observations (Section~\ref{sec:pol}).
We note that in-beam blending effects, particularly at 22\,GHz where the synthesized beam is larger, may affect the measured core spectral index. Extended jet emission within the beam could contribute additional flux to the core component, potentially flattening the observed spectrum. If the true core flux is lower than measured, the intrinsic spectral index could 
be even flatter ($\alpha \sim 0$), further supporting a scenario of sustained particle injection in a partially self-absorbed core region. Nevertheless, the consistency between the 22-43 and 43-86\,GHz pairs, which probe different spatial scales, suggests that beam blending is not the dominant effect shaping the observed flat spectrum.

\subsection{Faraday rotation: external screen vs. intrinsic jet effects}
\label{subsec:faraday_interpretation}

The exceptionally high rotation measure, $\mathrm{RM}=(7.81\pm0.16)\times10^{4}\ \mathrm{rad\,m^{-2}}$, significantly exceeds typical blazar-core values and requires interpretation. Two scenarios can explain this: an external Faraday screen or intrinsic jet opacity effects.
The spatial uniformity of the RM and lack of transverse gradients favor an external screen, likely associated with the Galactic interstellar medium. TXS~2013+370 lies close to the Galactic plane ($b \sim 2^{\circ}$) in the Cygnus region, where high column densities of magnetized plasma are well-documented \citep{refId0}. The $\sim50^{\circ}$ EVPA rotation between 43 and 86\,GHz is naturally explained by wavelength-dependent Faraday rotation 
($\chi \propto \lambda^2$). Internal jet processes would typically produce spatial RM gradients, which are not observed.
However, opacity effects cannot be excluded. The EVPA flip at 86\,GHz could reflect increasing transparency in the jet base, revealing intrinsic EVPA variations from jet curvature and helical magnetic field geometry \citep{Gabuzda2004,ONeill2019}, rather than true Faraday rotation. In this scenario, frequency-dependent opacity would cause the observed $\lambda^2$ EVPA behavior without actual rotation measure.
While both scenarios remain viable, the RM uniformity and the established Galactic foreground along this sight line \citep{refId0} favor the external screen interpretation. Multi-epoch polarimetric observations are needed to distinguish between these scenarios: external screens produce stable RM values, while opacity effects should vary with jet activity.

\section{Summary and conclusions}
\label{sec:conclusions}

In this work, we conducted polarimetric VLBI observations of \object{TXS~2013+370} at 22, 43, and 86\,GHz during an exceptional GeV outburst on 11 February 2021, achieving angular resolutions down to $\sim$0.1\,mas. This represents the first multi-frequency polarimetric VLBI study of this source. Our main findings are as follows:

\begin{enumerate}
    \item Detection of a new jet component during enhanced activity. Our imaging revealed a curved jet with a newly emerged component N2 located $\sim$60\,$\mu$as from the core, associated with enhanced multi-wavelength activity. Spectral analysis showed flat indices ($\alpha \gtrsim -0.5$) indicating ongoing particle acceleration.
    
    \item Strong polarization and exceptionally high Faraday rotation. Polarimetric observations revealed strong coherent polarization (3.3-4.5\%) and an exceptionally high Faraday rotation measure of $(7.81 \pm 0.16) \times 10^4$\,rad\,m$^{-2}$, indicating propagation through a dense, magnetized external screen, with internal jet opacity effects cannot be excluded.
    
    \item Precise localization of the $\gamma$-ray emission site. Cross-correlation analysis revealed a highly significant $\gamma$-ray lead of $(102 \pm 12)$\,days over 15\,GHz emission, corresponding to a separation of $(2.71 \pm 0.47)$\,pc. Combined with the core-apex distance constraint from \citet{refId0}, we locate the $\gamma$-ray emission at $(-0.66 \pm 1.08)$\,pc, with a large uncertainty that spans from the jet apex to $\sim0.42$ pc, consistent with sub-parsec scales favoring both the BLR and the dusty torus environments.
    
    \item Stable $\gamma$-ray production region despite variable delays. While the 2009 and 2021 flares show different $\gamma$-ray to radio delays, both locate the $\gamma$-ray emission to the same sub-parsec/parsec region, indicating that lag variations reflect changing opacity conditions rather than a moving dissipation site.
    
    \item Support for external Compton scattering models. Our results demonstrate that $\gamma$-ray production occurs in a spatially stable region in \object{TXS~2013+370}, anchored to specific jet features, challenging simple one-zone models and supporting scenarios where EC scattering on BLR or/and dusty torus photons dominate the high-energy emission.
\end{enumerate}

\begin{acknowledgements}

I would like to especially thank Prof. P. Papadopoulos for his invaluable contributions, mentoring, and support throughout this work. His lectures and discussions made the analysis and the paper writing an inspiring and rewarding journey of discovery. 

We also thank the anonymous referee for the valuable comments and suggestions, which greatly improved the paper.

This research has made use of data from the OVRO 40-m monitoring program (Richards, J. L. et al. 2011, ApJS, 194, 29), which is supported in part by NASA grants NNX08AW31G, NNX11A043G, and NNX14AQ89G and NSF grants AST-0808050 and AST-1109911.

The Submillimeter Array is a joint project between the Smithsonian Astrophysical Observatory and the Academia Sinica Institute of Astronomy and Astrophysics and is funded by the Smithsonian Institution and the Academia Sinica.

This research has made use of NASA's Astrophysics Data System.

The \textit{Fermi}-LAT Collaboration acknowledges the generous ongoing support from a number of agencies and institutes that have supported both the development and the operation of the LAT, as well as scientific data analysis. These include the National Aeronautics and Space Administration and the Department of Energy in the United States; the Commissariat \'a l’Energie Atomique and the Centre National de la Recherche Scientifique/Institut National de Physique Nucl\'eaire et de Physique des Particules in France; the Agenzia Spaziale Italiana and the Istituto Nazionale di Fisica Nucleare in Italy; the Ministry of Education, Culture, Sports, Science and Technology (MEXT), High Energy Accelerator Research Organization (KEK), and Japan Aerospace Exploration Agency (JAXA) in Japan; and the K. A. Wallenberg Foundation, the Swedish Research  Council, and  the  Swedish  National  Space  Board in Sweden. Additional support for science analysis during the operations phase is gratefully acknowledged from the Istituto Nazionale di Astrofisica in Italy and the Centre National d'Etudes Spatiales in France. This work was performed in part under DOE Contract DE-AC02-76SF00515.

We recognize that Maunakea is a culturally important site for the indigenous Hawaiian people; we are privileged to study the cosmos from its summit.

This work was supported by the Max Planck Institute for Radioastronomy (MPIfR) - Mexico Max Planck Partner Group led by V.M.P.-A.

This publication includes data based on observations with the 100-m telescope of the MPIfR at Effelsberg.

\end{acknowledgements}

\bibliographystyle{aa} 
\bibliography{aanda}

@article{1988ApJ...333..646E,
 adsurl = {http://adsabs.harvard.edu/abs/1988ApJ...333..646E},
 author = {{Edelson}, R.~A. and {Krolik}, J.~H.},
 doi = {10.1086/166773},
 journal = {\apj},
 month = {October},
 pages = {646-659},
 title = {{The discrete correlation function - A new method for analyzing unevenly sampled variability data}},
 volume = {333},
 year = {1988}
}

@article{1992A&A...256L..27D,
 adsurl = {http://adsabs.harvard.edu/abs/1992A%26A...256L..27D},
 author = {{Dermer}, C.~D. and {Schlickeiser}, R. and {Mastichiadis}, A.
},
 journal = {\aap},
 month = {March},
 pages = {L27-L30},
 title = {{High-energy gamma radiation from extragalactic radio sources}},
 volume = {256},
 year = {1992}
}

@article{1993ApJ...416..458D,
 adsurl = {http://adsabs.harvard.edu/abs/1993ApJ...416..458D},
 author = {{Dermer}, C.~D. and {Schlickeiser}, R.},
 doi = {10.1086/173251},
 journal = {\apj},
 month = {October},
 pages = {458},
 title = {{Model for the High-Energy Emission from Blazars}},
 volume = {416},
 year = {1993}
}

@article{1995PASP..107..803U,
 adsurl = {http://adsabs.harvard.edu/abs/1995PASP..107..803U},
 author = {{Urry}, C.~M. and {Padovani}, P.},
 doi = {10.1086/133630},
 eprint = {astro-ph/9506063},
 journal = {\pasp},
 month = {September},
 pages = {803},
 title = {{Unified Schemes for Radio-Loud Active Galactic Nuclei}},
 volume = {107},
 year = {1995}
}

@article{1999ApJ...514..138K,
 adsurl = {http://adsabs.harvard.edu/abs/1999ApJ...514..138K},
 author = {{Kataoka}, J. and {Mattox}, J.~R. and {Quinn}, J. and {Kubo}, H. and 
{Makino}, F. and {Takahashi}, T. and {Inoue}, S. and {Hartman}, R.~C. and 
{Madejski}, G.~M. and {Sreekumar}, P. and {Wagner}, S.~J.},
 doi = {10.1086/306918},
 eprint = {astro-ph/9811014},
 journal = {\apj},
 month = {March},
 pages = {138-147},
 title = {{High-Energy Emission from the TEV Blazar Markarian 501 during Multiwavelength Observations in 1996}},
 volume = {514},
 year = {1999}
}

@article{2000ApJ...542..740M,
 adsurl = {http://adsabs.harvard.edu/abs/2000ApJ...542..740M},
 author = {{Mukherjee}, R. and {Gotthelf}, E.~V. and {Halpern}, J. and 
{Tavani}, M.},
 doi = {10.1086/317054},
 eprint = {astro-ph/0005491},
 journal = {\apj},
 month = {October},
 pages = {740-749},
 title = {{Multiwavelength Examination of the COS B Field 2CG 075+00 Yields a Blazar Identification for 3EG J2016+3657}},
 volume = {542},
 year = {2000}
}

@article{2001ApJ...551.1016H,
 adsurl = {http://adsabs.harvard.edu/abs/2001ApJ...551.1016H},
 author = {{Halpern}, J.~P. and {Eracleous}, M. and {Mukherjee}, R. and 
{Gotthelf}, E.~V.},
 doi = {10.1086/320238},
 eprint = {astro-ph/0012508},
 journal = {\apj},
 month = {April},
 pages = {1016-1023},
 title = {{3EG J2016+3657: Confirming an EGRET Blazar behind the Galactic Plane}},
 volume = {551},
 year = {2001}
}

@article{2001ApJ...556..738J,
 adsurl = {http://adsabs.harvard.edu/abs/2001ApJ...556..738J},
 author = {{Jorstad}, S.~G. and {Marscher}, A.~P. and {Mattox}, J.~R. and 
{Aller}, M.~F. and {Aller}, H.~D. and {Wehrle}, A.~E. and {Bloom}, S.~D.
},
 doi = {10.1086/321605},
 eprint = {astro-ph/0102012},
 journal = {\apj},
 month = {August},
 pages = {738-748},
 title = {{Multiepoch Very Long Baseline Array Observations of EGRET-detected Quasars and BL Lacertae Objects: Connection between Superluminal Ejections and Gamma-Ray Flares in Blazars}},
 volume = {556},
 year = {2001}
}

@article{Gabuzda2004,
  author  = {Gabuzda, Denise C. and Pushkarev, Alexander B. and Cawthorne, Timothy V.},
  title   = {Analysis of $\lambda = 6$ cm VLBI polarization observations of a complete sample of northern BL Lacertae objects},
  journal = {MNRAS},
  year    = {2004},
  volume  = {350},
  number  = {2},
  pages   = {529--549},
  doi     = {10.1111/j.1365-2966.2004.07654.x}
}

@article{2012ApJ...746..159K,
 adsurl = {http://adsabs.harvard.edu/abs/2012ApJ...746..159K},
 archiveprefix = {arXiv},
 author = {{Kara}, E. and {Errando}, M. and {Max-Moerbeck}, W. and {Aliu}, E. and 
{B{\"o}ttcher}, M. and {Fortin}, P. and {Halpern}, J.~P. and 
{Mukherjee}, R. and {Readhead}, A.~C.~S. and {Richards}, J.~L.
},
 doi = {10.1088/0004-637X/746/2/159},
 eid = {159},
 eprint = {1112.3312},
 journal = {ApJ},
 month = {February},
 pages = {159},
 primaryclass = {astro-ph.HE},
 title = {{Gamma-Ray Emission from Two Blazars Behind the Galactic Plane: B2013+370 and B2023+336}},
 volume = {746},
 year = {2012}
}

@article{2012ApJ...758L..15D,
 adsurl = {http://adsabs.harvard.edu/abs/2012ApJ...758L..15D},
 archiveprefix = {arXiv},
 author = {{Dotson}, A. and {Georganopoulos}, M. and {Kazanas}, D. and 
{Perlman}, E.~S.},
 doi = {10.1088/2041-8205/758/1/L15},
 eid = {L15},
 eprint = {1209.2053},
 journal = {\apjl},
 month = {October},
 pages = {L15},
 primaryclass = {astro-ph.HE},
 title = {{A Method for Localizing Energy Dissipation in Blazars Using Fermi Variability}},
 volume = {758},
 year = {2012}
}

@article{2013ApJ...764..135S,
 adsurl = {http://adsabs.harvard.edu/abs/2013ApJ...764..135S},
 archiveprefix = {arXiv},
 author = {{Shaw}, M.~S. and {Romani}, R.~W. and {Cotter}, G. and {Healey}, S.~E. and 
{Michelson}, P.~F. and {Readhead}, A.~C.~S. and {Richards}, J.~L. and 
{Max-Moerbeck}, W. and {King}, O.~G. and {Potter}, W.~J.},
 doi = {10.1088/0004-637X/764/2/135},
 eid = {135},
 eprint = {1301.0323},
 journal = {\apj},
 month = {February},
 pages = {135},
 primaryclass = {astro-ph.HE},
 title = {{Spectroscopy of the Largest Ever {$\gamma$}-Ray-selected BL Lac Sample}},
 volume = {764},
 year = {2013}
}

@article{2013MNRAS.433..907E,
 adsurl = {http://adsabs.harvard.edu/abs/2013MNRAS.433..907E},
 archiveprefix = {arXiv},
 author = {{Emmanoulopoulos}, D. and {McHardy}, I.~M. and {Papadakis}, I.~E.
},
 doi = {10.1093/mnras/stt764},
 eprint = {1305.0304},
 journal = {\mnras},
 month = {August},
 pages = {907-927},
 primaryclass = {astro-ph.IM},
 title = {{Generating artificial light curves: revisited and updated}},
 volume = {433},
 year = {2013}
}

@article{2015MNRAS.448.1060G,
 adsurl = {http://adsabs.harvard.edu/abs/2015MNRAS.448.1060G},
 archiveprefix = {arXiv},
 author = {{Ghisellini}, G. and {Tavecchio}, F.},
 doi = {10.1093/mnras/stv055},
 eprint = {1501.03504},
 journal = {\mnras},
 month = {April},
 pages = {1060-1077},
 primaryclass = {astro-ph.HE},
 title = {{Fermi/LAT broad emission line blazars}},
 volume = {448},
 year = {2015}
}

@article{2016AJ....152...12L,
 adsurl = {http://adsabs.harvard.edu/abs/2016AJ....152...12L},
 archiveprefix = {arXiv},
 author = {{Lister}, M.~L. and {Aller}, M.~F. and {Aller}, H.~D. and {Homan}, D.~C. and 
{Kellermann}, K.~I. and {Kovalev}, Y.~Y. and {Pushkarev}, A.~B. and 
{Richards}, J.~L. and {Ros}, E. and {Savolainen}, T.},
 doi = {10.3847/0004-6256/152/1/12},
 eid = {12},
 eprint = {1603.03882},
 journal = {\aj},
 month = {July},
 pages = {12},
 title = {{MOJAVE: XIII. Parsec-scale AGN Jet Kinematics Analysis Based on 19 years of VLBA Observations at 15 GHz}},
 volume = {152},
 year = {2016}
}

@article{2017A&ARv..25....4B,
 adsurl = {http://adsabs.harvard.edu/abs/2017A%26ARv..25....4B},
 archiveprefix = {arXiv},
 author = {{Boccardi}, B. and {Krichbaum}, T.~P. and {Ros}, E. and {Zensus}, J.~A.
},
 doi = {10.1007/s00159-017-0105-6},
 eid = {4},
 eprint = {1711.07548},
 journal = {\aapr},
 month = {November},
 pages = {4},
 primaryclass = {astro-ph.HE},
 title = {{Radio observations of active galactic nuclei with mm-VLBI}},
 volume = {25},
 year = {2017}
}

@article{2017ApJ...846...98J,
 adsurl = {http://adsabs.harvard.edu/abs/2017ApJ...846...98J},
 archiveprefix = {arXiv},
 author = {{Jorstad}, S.~G. and {Marscher}, A.~P. and {Morozova}, D.~A. and 
{Troitsky}, I.~S. and {Agudo}, I. and {Casadio}, C. and {Foord}, A. and 
{G{\'o}mez}, J.~L. and {MacDonald}, N.~R. and {Molina}, S.~N. and 
{L{\"a}hteenm{\"a}ki}, A. and {Tammi}, J. and {Tornikoski}, M.
},
 doi = {10.3847/1538-4357/aa8407},
 eid = {98},
 eprint = {1711.03983},
 journal = {\apj},
 month = {September},
 pages = {98},
 title = {{Kinematics of Parsec-scale Jets of Gamma-Ray Blazars at 43 GHz within the VLBA-BU-BLAZAR Program}},
 volume = {846},
 year = {2017}
}

@ARTICLE{2008MNRAS.385..283C,
       author = {{Celotti}, Annalisa and {Ghisellini}, Gabriele},
        title = "{The power of blazar jets}",
      journal = {\mnras},
         year = "2008",
        month = "Mar",
       volume = {385},
       number = {1},
        pages = {283-300},
          doi = {10.1111/j.1365-2966.2007.12758.x},
archivePrefix = {arXiv},
       eprint = {0711.4112},
 primaryClass = {astro-ph},
       adsurl = {https://ui.adsabs.harvard.edu/abs/2008MNRAS.385..283C}
}

@ARTICLE{1992ApJ...397L...5M,
       author = {{Maraschi}, L. and {Ghisellini}, G. and {Celotti}, A.},
        title = "{A Jet Model for the Gamma-Ray--emitting Blazar 3C 279}",
      journal = {\apjl},
         year = "1992",
        month = "Sep",
       volume = {397},
        pages = {L5},
          doi = {10.1086/186531},
       adsurl = {https://ui.adsabs.harvard.edu/abs/1992ApJ...397L...5M}
}

@ARTICLE{2000ApJ...545..107B,
   author = {{B{\l}a{\.z}ejowski}, M. and {Sikora}, M. and {Moderski}, R. and 
	{Madejski}, G.~M.},
    title = "{Comptonization of Infrared Radiation from Hot Dust by Relativistic Jets in Quasars}",
  journal = {\apj},
   eprint = {astro-ph/0008154},
     year = 2000,
    month = dec,
   volume = 545,
    pages = {107-116},
      doi = {10.1086/317791},
   adsurl = {https://ui.adsabs.harvard.edu/abs/2000ApJ...545..107B}
}

@ARTICLE{2008MNRAS.386L..28G,
   author = {{Ghisellini}, G. and {Tavecchio}, F.},
    title = "{Rapid variability in TeV blazars: the case of PKS2155-304}",
  journal = {\mnras},
archivePrefix = "arXiv",
   eprint = {0801.2569},
     year = 2008,
    month = may,
   volume = 386,
    pages = {L28-L32},
      doi = {10.1111/j.1745-3933.2008.00454.x},
   adsurl = {https://ui.adsabs.harvard.edu/abs/2008MNRAS.386L..28G}
}

@ARTICLE{2009ApJ...692...32D,
   author = {{Dermer}, C.~D. and {Finke}, J.~D. and {Krug}, H. and {B{\"o}ttcher}, M.
	},
    title = "{Gamma-Ray Studies of Blazars: Synchro-Compton Analysis of Flat Spectrum Radio Quasars}",
  journal = {\apj},
archivePrefix = "arXiv",
   eprint = {0808.3185},
     year = 2009,
    month = feb,
   volume = 692,
    pages = {32-46},
      doi = {10.1088/0004-637X/692/1/32},
   adsurl = {https://ui.adsabs.harvard.edu/abs/2009ApJ...692...32D}
}

@ARTICLE{2014A&A...571L...2R,
       author = {{Rani}, B. and {Krichbaum}, T.~P. and {Marscher}, A.~P. and
         {Jorstad}, S.~G. and {Hodgson}, J.~A. and {Fuhrmann}, L. and
         {Zensus}, J.~A.},
        title = "{Jet outflow and gamma-ray emission correlations in S5 0716+714}",
      journal = {\aap},
         year = "2014",
        month = "Nov",
       volume = {571},
          eid = {L2},
        pages = {L2},
          doi = {10.1051/0004-6361/201424796},
archivePrefix = {arXiv},
       eprint = {1410.0196},
 primaryClass = {astro-ph.HE},
       adsurl = {https://ui.adsabs.harvard.edu/abs/2014A&A...571L...2R}
}

@INPROCEEDINGS{1994cers.conf..207L,
       author = {{Lepp{\"a}nen}, K.~J. and {Zensus}, J.~A. and {Diamond}, P.~D.},
        title = "{High Frequency Polarization Imaging with the VLBA}",
    booktitle = {Compact Extragalactic Radio Sources},
         year = 1994,
       editor = {{Zensus}, J. Anton and {Kellermann}, Kenneth I.},
        month = jan,
        pages = {207},
       adsurl = {https://ui.adsabs.harvard.edu/abs/1994cers.conf..207L}
}

@INPROCEEDINGS{1999ASPC..180..301F,
       author = {{Fomalont}, Ed B.},
        title = "{Image Analysis}",
    booktitle = {Synthesis Imaging in Radio Astronomy II},
         year = 1999,
       editor = {{Taylor}, G.~B. and {Carilli}, C.~L. and {Perley}, R.~A.},
       series = {ASP Conference Series},
       volume = {180},
        month = jan,
        pages = {301},
       adsurl = {https://ui.adsabs.harvard.edu/abs/1999ASPC..180..301F}
}

@ARTICLE{2005astro.ph..3225L,
       author = {{Lobanov}, A.~P.},
        title = "{Resolution limits in astronomical images}",
      journal = {arXiv e-prints},
         year = 2005,
        month = mar,
          eid = {astro-ph/0503225},
        pages = {astro-ph/0503225},
          doi = {10.48550/arXiv.astro-ph/0503225},
archivePrefix = {arXiv},
       eprint = {astro-ph/0503225},
 primaryClass = {astro-ph},
       adsurl = {https://ui.adsabs.harvard.edu/abs/2005astro.ph..3225L}
}

@ARTICLE{2012A&A...537A..70S,
       author = {{Schinzel}, F.~K. and {Lobanov}, A.~P. and {Taylor}, G.~B. and {Jorstad}, S.~G. and {Marscher}, A.~P. and {Zensus}, J.~A.},
        title = "{Relativistic outflow drives {\ensuremath{\gamma}}-ray emission in 3C 345}",
      journal = {\aap},
         year = 2012,
        month = jan,
       volume = {537},
          eid = {A70},
        pages = {A70},
          doi = {10.1051/0004-6361/201117705},
archivePrefix = {arXiv},
       eprint = {1111.2045},
 primaryClass = {astro-ph.CO},
       adsurl = {https://ui.adsabs.harvard.edu/abs/2012A&A...537A..70S}
}

@ARTICLE{2024A&A...682A.154T,
       author = {{Traianou}, Efthalia and {Krichbaum}, Thomas P. and {G{\'o}mez}, Jos{\'e} L. and {Lico}, Rocco and {Paraschos}, Georgios Filippos and {Cho}, Ilje and {Ros}, Eduardo and {Zhao}, Guang-Yao and {Liodakis}, Ioannis and {Dahale}, Rohan and {Toscano}, Teresa and {Fuentes}, Antonio and {Foschi}, Marianna and {Casadio}, Carolina and {MacDonald}, Nicholas and {Kim}, Jae-Young and {Hervet}, Olivier and {Jorstad}, Svetlana and {Lobanov}, Andrei P. and {Hodgson}, Jeffrey and {Myserlis}, Ioannis and {Agudo}, Ivan and {Zensus}, Anton J. and {Marscher}, Alan P.},
        title = "{Lost in the curve: Investigating the disappearing knots in blazar 3C 454.3}",
      journal = {\aap},
         year = 2024,
        month = feb,
       volume = {682},
          eid = {A154},
        pages = {A154},
          doi = {10.1051/0004-6361/202347267},
archivePrefix = {arXiv},
       eprint = {2312.15556},
 primaryClass = {astro-ph.HE},
       adsurl = {https://ui.adsabs.harvard.edu/abs/2024A&A...682A.154T}
}

@article{Ulrich_1997,
  author={Ulrich, M.-H. and Maraschi, L. and Urry, C. M.},
  title={Variability of Active Galactic Nuclei},
  journal={ARA\&A},
  year={1997},
  volume={35},
  pages={445--502},
  doi={10.1146/annurev.astro.35.1.445}
}

@ARTICLE{2009ApJ...697.1071A,
       author = {{Atwood}, W.~B. and {Abdo}, A.~A. and {Ackermann}, M. and {Althouse}, W. and {Anderson}, B. and {Axelsson}, M. and {Baldini}, L. and {Ballet}, J. and {Band}, D.~L. and {Barbiellini}, G. and {Bartelt}, J. and {Bastieri}, D. and {Baughman}, B.~M. and {Bechtol}, K. and {B{\'e}d{\'e}r{\`e}de}, D. and {Bellardi}, F. and {Bellazzini}, R. and {Berenji}, B. and {Bignami}, G.~F. and {Bisello}, D. and {Bissaldi}, E. and {Blandford}, R.~D. and {Bloom}, E.~D. and {Bogart}, J.~R. and {Bonamente}, E. and {Bonnell}, J. and {Borgland}, A.~W. and {Bouvier}, A. and {Bregeon}, J. and {Brez}, A. and {Brigida}, M. and {Bruel}, P. and {Burnett}, T.~H. and {Busetto}, G. and {Caliandro}, G.~A. and {Cameron}, R.~A. and {Caraveo}, P.~A. and {Carius}, S. and {Carlson}, P. and {Casandjian}, J.~M. and {Cavazzuti}, E. and {Ceccanti}, M. and {Cecchi}, C. and {Charles}, E. and {Chekhtman}, A. and {Cheung}, C.~C. and {Chiang}, J. and {Chipaux}, R. and {Cillis}, A.~N. and {Ciprini}, S. and {Claus}, R. and {Cohen-Tanugi}, J. and {Condamoor}, S. and {Conrad}, J. and {Corbet}, R. and {Corucci}, L. and {Costamante}, L. and {Cutini}, S. and {Davis}, D.~S. and {Decotigny}, D. and {DeKlotz}, M. and {Dermer}, C.~D. and {de Angelis}, A. and {Digel}, S.~W. and {do Couto e Silva}, E. and {Drell}, P.~S. and {Dubois}, R. and {Dumora}, D. and {Edmonds}, Y. and {Fabiani}, D. and {Farnier}, C. and {Favuzzi}, C. and {Flath}, D.~L. and {Fleury}, P. and {Focke}, W.~B. and {Funk}, S. and {Fusco}, P. and {Gargano}, F. and {Gasparrini}, D. and {Gehrels}, N. and {Gentit}, F. -X. and {Germani}, S. and {Giebels}, B. and {Giglietto}, N. and {Giommi}, P. and {Giordano}, F. and {Glanzman}, T. and {Godfrey}, G. and {Grenier}, I.~A. and {Grondin}, M. -H. and {Grove}, J.~E. and {Guillemot}, L. and {Guiriec}, S. and {Haller}, G. and {Harding}, A.~K. and {Hart}, P.~A. and {Hays}, E. and {Healey}, S.~E. and {Hirayama}, M. and {Hjalmarsdotter}, L. and {Horn}, R. and {Hughes}, R.~E. and {J{\'o}hannesson}, G. and {Johansson}, G. and {Johnson}, A.~S. and {Johnson}, R.~P. and {Johnson}, T.~J. and {Johnson}, W.~N. and {Kamae}, T. and {Katagiri}, H. and {Kataoka}, J. and {Kavelaars}, A. and {Kawai}, N. and {Kelly}, H. and {Kerr}, M. and {Klamra}, W. and {Kn{\"o}dlseder}, J. and {Kocian}, M.~L. and {Komin}, N. and {Kuehn}, F. and {Kuss}, M. and {Landriu}, D. and {Latronico}, L. and {Lee}, B. and {Lee}, S. -H. and {Lemoine-Goumard}, M. and {Lionetto}, A.~M. and {Longo}, F. and {Loparco}, F. and {Lott}, B. and {Lovellette}, M.~N. and {Lubrano}, P. and {Madejski}, G.~M. and {Makeev}, A. and {Marangelli}, B. and {Massai}, M.~M. and {Mazziotta}, M.~N. and {McEnery}, J.~E. and {Menon}, N. and {Meurer}, C. and {Michelson}, P.~F. and {Minuti}, M. and {Mirizzi}, N. and {Mitthumsiri}, W. and {Mizuno}, T. and {Moiseev}, A.~A. and {Monte}, C. and {Monzani}, M.~E. and {Moretti}, E. and {Morselli}, A. and {Moskalenko}, I.~V. and {Murgia}, S. and {Nakamori}, T. and {Nishino}, S. and {Nolan}, P.~L. and {Norris}, J.~P. and {Nuss}, E. and {Ohno}, M. and {Ohsugi}, T. and {Omodei}, N. and {Orlando}, E. and {Ormes}, J.~F. and {Paccagnella}, A. and {Paneque}, D. and {Panetta}, J.~H. and {Parent}, D. and {Pearce}, M. and {Pepe}, M. and {Perazzo}, A. and {Pesce-Rollins}, M. and {Picozza}, P. and {Pieri}, L. and {Pinchera}, M. and {Piron}, F. and {Porter}, T.~A. and {Poupard}, L. and {Rain{\`o}}, S. and {Rando}, R. and {Rapposelli}, E. and {Razzano}, M. and {Reimer}, A. and {Reimer}, O. and {Reposeur}, T. and {Reyes}, L.~C. and {Ritz}, S. and {Rochester}, L.~S. and {Rodriguez}, A.~Y. and {Romani}, R.~W. and {Roth}, M. and {Russell}, J.~J. and {Ryde}, F. and {Sabatini}, S. and {Sadrozinski}, H.~F. -W. and {Sanchez}, D. and {Sander}, A. and {Sapozhnikov}, L. and {Parkinson}, P.~M. Saz and {Scargle}, J.~D. and {Schalk}, T.~L. and {Scolieri}, G.},
        title = "{The Large Area Telescope on the Fermi Gamma-Ray Space Telescope Mission}",
      journal = {\apj},
         year = 2009,
        month = jun,
       volume = {697},
       number = {2},
        pages = {1071-1102},
          doi = {10.1088/0004-637X/697/2/1071},
archivePrefix = {arXiv},
       eprint = {0902.1089},
 primaryClass = {astro-ph.IM},
       adsurl = {https://ui.adsabs.harvard.edu/abs/2009ApJ...697.1071A}
}

@ARTICLE{2025A&A...700A..16T,
       author = {{Traianou}, E. and {G{\'o}mez}, J.~L. and {Cho}, I. and {Chael}, A. and {Fuentes}, A. and {Myserlis}, I. and {Wielgus}, M. and {Zhao}, G. -Y. and {Lico}, R. and {Moriyama}, K. and {Dey}, L. and {Bruni}, G. and {Dahale}, R. and {Toscano}, T. and {Gurvits}, L.~I. and {Lisakov}, M.~M. and {Kovalev}, Y.~Y. and {Lobanov}, A.~P. and {Pushkarev}, A.~B. and {Sokolovsky}, K.~V.},
        title = "{Revealing a ribbon-like jet in OJ 287 with RadioAstron}",
      journal = {\aap},
         year = 2025,
        month = aug,
       volume = {700},
          eid = {A16},
        pages = {A16},
          doi = {10.1051/0004-6361/202554929},
archivePrefix = {arXiv},
       eprint = {2508.01747},
 primaryClass = {astro-ph.HE},
       adsurl = {https://ui.adsabs.harvard.edu/abs/2025A&A...700A..16T}
}

@ARTICLE{2006ApJ...641..689V,
       author = {{Vestergaard}, Marianne and {Peterson}, Bradley M.},
        title = "{Determining Central Black Hole Masses in Distant Active Galaxies and Quasars. II. Improved Optical and UV Scaling Relationships}",
      journal = {\apj},
         year = 2006,
        month = apr,
       volume = {641},
       number = {2},
        pages = {689-709},
          doi = {10.1086/500572},
archivePrefix = {arXiv},
       eprint = {astro-ph/0601303},
 primaryClass = {astro-ph},
       adsurl = {https://ui.adsabs.harvard.edu/abs/2006ApJ...641..689V}
}

@article{McKinney_2009,
  author={McKinney, J. C. and Blandford, R. D.},
  title={Stability of relativistic jets from rotating, accreting black holes via fully three-dimensional magnetohydrodynamic simulations},
  journal={MNRAS},
  year={2009},
  volume={394},
  pages={L126--L130},
  doi={10.1111/j.1745-3933.2009.00625.x}
}

@article{Lyutikov_2013,
  author={Lyutikov, M.},
  title={Magnetic field structure of relativistic jets without current sheets},
  journal={MNRAS},
  year={2012},
  volume={419},
  pages={3048--3059},
  doi={10.1111/j.1365-2966.2011.19946.x}
}

@article{10.1093/mnras/stx2435,
    author = {Agudo, Iván and Thum, Clemens and Molina, Sol N and Casadio, Carolina and Wiesemeyer, Helmut and Morris, David and Paubert, Gabriel and Gómez, José L and Kramer, Carsten},
    title = {POLAMI: Polarimetric Monitoring of AGN at Millimetre Wavelengths – I. The programme, calibration and calibrator data products},
    journal = {MNRAS},
    volume = {474},
    number = {2},
    pages = {1427-1435},
    year = {2017},
    month = {09},
    issn = {0035-8711},
    doi = {10.1093/mnras/stx2435},
    url = {https://doi.org/10.1093/mnras/stx2435},
    eprint = {https://academic.oup.com/mnras/article-pdf/474/2/1427/24011516/stx2435.pdf},
}

@article{Sironi_2016,
  author={Sironi, L. and Petropoulou, M. and Giannios, D.},
  title={Relativistic jets shine through shocks or magnetic reconnection?},
      journal={MNRAS},
  year={2015},
  volume={450},
  pages={183--191},
  doi={10.1093/mnras/stv641}
}

@inproceedings{1997ASPC..125...77S,
 adsurl = {https://ui.adsabs.harvard.edu/abs/1997ASPC..125...77S},
 author = {{Shepherd}, M.~C.},
 booktitle = {ADASS VI},
 editor = {{Hunt}, Gareth and {Payne}, Harry},
 month = {January},
 pages = {77},
 series = {ASP Conference Series},
 title = {{Difmap: an Interactive Program for Synthesis Imaging}},
 volume = {125},
 year = {1997}
}

@inproceedings{1990apaa.conf..125G,
 adsurl = {http://adsabs.harvard.edu/abs/1990apaa.conf..125G},
 author = {{Greisen}, E.~W.},
 booktitle = {Acquisition, Processing and Archiving of Astronomical Images},
 editor = {{Longo}, G. and {Sedmak}, G.},
 pages = {125-142},
 title = {{The Astronomical Image Processing System.}},
 year = {1990}
}

@article{ refId0,
	author = {{Traianou, E.} and {Krichbaum, T. P.} and {Boccardi, B.} and {Angioni, R.} and {Rani, B.} and {Liu, J.} and {Ros, E.} and {Bach, U.} and {Sokolovsky, K. V.} and {Lisakov, M. M.} and {Kiehlmann, S.} and {Gurwell, M.} and {Zensus, J. A.}},
	title = {Localizing the $\gamma$-ray emitting region in the blazar TXS 2013+370},
	journal = {A$\&$A},
	year = 2020,
	volume = 634,
	pages = "A112",
}

@article{Kara_2012,
year = {2012},
month = {feb},
publisher = {The American Astronomical Society},
volume = {746},
number = {2},
pages = {159},
author = {Kara, E. and Errando, M. and Max-Moerbeck, W. and Aliu, E. and Böttcher, M. and Fortin, P. and Halpern, J. P. and Mukherjee, R. and Readhead, A. C. S. and Richards, J. L.},
title = {GAMMA-RAY EMISSION FROM TWO BLAZARS BEHIND THE GALACTIC PLANE: B2013+370 AND B2023+336},
journal = {ApJ},

}

@article{Hovatta_2012,
doi = {10.1088/0004-6256/144/4/105},
url = {https://dx.doi.org/10.1088/0004-6256/144/4/105},
year = {2012},
month = {sep},
publisher = {The American Astronomical Society},
volume = {144},
number = {4},
pages = {105},
author = {Hovatta, Talvikki and Lister, Matthew L. and Aller, Margo F. and Aller, Hugh D. and Homan, Daniel C. and Kovalev, Yuri Y. and Pushkarev, Alexander B. and Savolainen, Tuomas},
title = {MOJAVE: MONITORING OF JETS IN ACTIVE GALACTIC NUCLEI WITH VLBA EXPERIMENTS. VIII. FARADAY ROTATION IN PARSEC-SCALE AGN JETS},
journal = {ApJ},
}

@article{Gomez_2022,
doi = {10.3847/1538-4357/ac3bcc},
url = {https://dx.doi.org/10.3847/1538-4357/ac3bcc},
year = {2022},
month = {jan},
publisher = {The American Astronomical Society},
volume = {924},
number = {2},
pages = {122},
author = {Gómez, José L. and Traianou, Efthalia and Krichbaum, Thomas P. and Lobanov, Andrei P. and Fuentes, Antonio and Lico, Rocco and Zhao, Guang-Yao and Bruni, Gabriele and Kovalev, Yuri Y. and Lähteenmäki, Anne and Voitsik, Petr A. and Lisakov, Mikhail M. and Angelakis, Emmanouil and Bach, Uwe and Casadio, Carolina and Cho, Ilje and Dey, Lankeswar and Gopakumar, Achamveedu and Gurvits, Leonid I. and Jorstad, Svetlana and Kovalev, Yuri A. and Lister, Matthew L. and Marscher, Alan P. and Myserlis, Ioannis and Pushkarev, Alexander B. and Ros, Eduardo and Savolainen, Tuomas and Tornikoski, Merja and Valtonen, Mauri J. and Zensus, Anton},
title = {Probing the Innermost Regions of AGN Jets and Their Magnetic Fields with RadioAstron. V. Space and Ground Millimeter-VLBI Imaging of OJ 287},
journal = {ApJ},
}

@article{bartolini2025,
      author={V. Bartolini and D. Dallacasa and J. L. Gómez and M. Giroletti and R. Lico and J. D. Livingston},
	title = {Multifrequency simultaneous VLBA view of the radio source 3C 111},
	DOI= "10.1051/0004-6361/202449878",
	url= "https://doi.org/10.1051/0004-6361/202449878",
	journal = {A$\&$A},
	year = 2025,
	volume = 698,
	pages = "A123",
}

@article{Pushkarev_2010,

year = {2010},
month = {sep},
publisher = {The American Astronomical Society},
volume = {722},
number = {1},
pages = {L7},
author = {Pushkarev, A. B. and Kovalev, Y. Y. and Lister, M. L.},
title = {RADIO/GAMMA-RAY TIME DELAY IN THE PARSEC-SCALE CORES OF ACTIVE GALACTIC NUCLEI},
journal = {ApJL},

}

@article{Cavaliere_2002,
doi = {10.1086/339778},
url = {https://dx.doi.org/10.1086/339778},
year = {2002},
month = {may},
publisher = {},
volume = {571},
number = {1},
pages = {226},
author = {Cavaliere, A. and D’Elia, V.},
title = {The Blazar Main Sequence},
journal = {ApJ}
}

@inproceedings{Sikora_2001,
   title={Blazars},
   volume={558},
   ISSN={0094-243X},
   url={http://dx.doi.org/10.1063/1.1370797},
   DOI={10.1063/1.1370797},
   booktitle={AIP Conference Proceedings},
   publisher={AIP},
   author={Sikora, Marek},
   year={2001},
   pages={275–288} }

@ARTICLE{2023A&A...669A..32P,
       author = {{Paraschos}, G.~F. and {Mpisketzis}, V. and {Kim}, J. -Y. and {Witzel}, G. and {Krichbaum}, T.~P. and {Zensus}, J.~A. and {Gurwell}, M.~A. and {L{\"a}hteenm{\"a}ki}, A. and {Tornikoski}, M. and {Kiehlmann}, S. and {Readhead}, A.~C.~S.},
        title = "{A multi-band study and exploration of the radio wave-{\ensuremath{\gamma}}-ray connection in 3C 84}",
      journal = {\aap},
         year = 2023,
        month = jan,
       volume = {669},
          eid = {A32},
        pages = {A32},
          doi = {10.1051/0004-6361/202244814},
archivePrefix = {arXiv},
       eprint = {2210.09795},
 primaryClass = {astro-ph.HE},
       adsurl = {https://ui.adsabs.harvard.edu/abs/2023A&A...669A..32P}
}

@ARTICLE{2020ApJ...891...68C,
       author = {{Chavushyan}, Vahram and {Pati{\~n}o-{\'A}lvarez}, Victor M. and {Amaya-Almaz{\'a}n}, Ra{\'u}l A. and {Carrasco}, Luis},
        title = "{Flare-like Variability of the Mg II {\ensuremath{\lambda}}2798 {\r{A}} Emission Line and UV Fe II Band in the Blazar CTA 102}",
      journal = {\apj},
         year = 2020,
        month = mar,
       volume = {891},
       number = {1},
          eid = {68},
        pages = {68},
          doi = {10.3847/1538-4357/ab6ef6},
archivePrefix = {arXiv},
       eprint = {2001.08296},
 primaryClass = {astro-ph.HE},
       adsurl = {https://ui.adsabs.harvard.edu/abs/2020ApJ...891...68C}
}

@article{ONeill2019,
  author  = {O'Neill, Shane and Kiehlmann, Sebastian and Readhead, A. C. S. and Pearson, T. J. and Carini, M. T. and others},
  title   = {Investigating the multiwavelength behaviour of the flat spectrum radio quasar CTA 102 during 2013--2017},
  journal = {MNRAS},
  year    = {2019},
  volume  = {490},
  number  = {2},
  pages   = {1961--1976},
  doi     = {10.1093/mnras/stz2750}
}

@article{Marscher2008,
  author  = {Marscher, Alan P. and Jorstad, Svetlana G. and D'Arcangelo, Francine D. and Smith, Peter S. and Williams, Gloria G. and Larionov, Valeri M. and Oh, H. and Olmstead, Alice R. and Aller, Margo F. and Aller, Hugh D. and McHardy, Ian M. and Lähteenmäki, Anne and Tornikoski, Merja and Valtaoja, Esko and Hagen-Thorn, Vera A. and Kopatskaya, Evgenia N. and Gear, Walter K. and Tosti, Giorgio and Kurtanidze, Omar M. and Nikolashvili, Mihail G. and Sigua, L. A. and Miller, Hillary R.},
  title   = {The inner jet of an active galactic nucleus as revealed by a radio-to-$\gamma$-ray outburst},
  journal = {Nature},
  year    = {2008},
  volume  = {452},
  pages   = {966--969},
  doi     = {10.1038/nature06895}
}

@ARTICLE{2012AJ....144..105H,
       author = {{Hovatta}, Talvikki and {Lister}, Matthew L. and {Aller}, Margo F. and {Aller}, Hugh D. and {Homan}, Daniel C. and {Kovalev}, Yuri Y. and {Pushkarev}, Alexander B. and {Savolainen}, Tuomas},
        title = "{MOJAVE: Monitoring of Jets in Active Galactic Nuclei with VLBA Experiments. VIII. Faraday Rotation in Parsec-scale AGN Jets}",
      journal = {\aj},
         year = 2012,
        month = oct,
       volume = {144},
       number = {4},
          eid = {105},
        pages = {105},
          doi = {10.1088/0004-6256/144/4/105},
archivePrefix = {arXiv},
       eprint = {1205.6746},
 primaryClass = {astro-ph.CO},
       adsurl = {https://ui.adsabs.harvard.edu/abs/2012AJ....144..105H}
}

@ARTICLE{2013MNRAS.429.3551A,
       author = {{Algaba}, J.~C.},
        title = "{High-frequency very long baseline interferometry rotation measure of eight active galactic nuclei}",
      journal = {\mnras},
         year = 2013,
        month = mar,
       volume = {429},
       number = {4},
        pages = {3551-3563},
          doi = {10.1093/mnras/sts624},
archivePrefix = {arXiv},
       eprint = {1212.3423},
 primaryClass = {astro-ph.HE},
       adsurl = {https://ui.adsabs.harvard.edu/abs/2013MNRAS.429.3551A}
}

@article{2016A&A586A60,
  author  = {Karamanavis, V. and Fuhrmann, L. and Krichbaum, T. P. and Angelakis, E. and Hodgson, J. and Nestoras, I. and Myserlis, I. and Zensus, J. A. and Sievers, A. and Ciprini, S.},
  title   = {PKS 1502+106: A high-redshift Fermi blazar at extreme angular resolution. Structural dynamics with VLBI imaging up to 86 GHz},
  journal = {A\&A},
  year    = {2016},
  volume  = {586},
  pages   = {A60},
  doi     = {10.1051/0004-6361/201527225}
}

@INPROCEEDINGS{2016JPhCS.718e2032R,
       author = {{Rani}, B. and {Krichbaum}, T.~P. and {Hodgson}, J.~A. and {Zensus}, J.~A.},
        title = "{Location and origin of gamma-rays in blazars}",
    booktitle = {JPCS},
         year = 2016,
       series = {JPCS},
       volume = {718},
        month = may,
          eid = {052032},
        pages = {052032},
          doi = {10.1088/1742-6596/718/5/052032},
archivePrefix = {arXiv},
       eprint = {1601.04693},
 primaryClass = {astro-ph.HE},
       adsurl = {https://ui.adsabs.harvard.edu/abs/2016JPhCS.718e2032R}
}

@ARTICLE{1998PASP..110..660P,
       author = {{Peterson}, Bradley M. and {Wanders}, Ignaz and {Horne}, Keith and {Collier}, Stefan and {Alexander}, Tal and {Kaspi}, Shai and {Maoz}, Dan},
        title = "{On Uncertainties in Cross-Correlation Lags and the Reality of Wavelength-dependent Continuum Lags in Active Galactic Nuclei}",
      journal = {\pasp},
         year = 1998,
        month = jun,
       volume = {110},
       number = {748},
        pages = {660-670},
          doi = {10.1086/316177},
archivePrefix = {arXiv},
       eprint = {astro-ph/9802103},
 primaryClass = {astro-ph},
       adsurl = {https://ui.adsabs.harvard.edu/abs/1998PASP..110..660P}
}

@article{Sironi_2014,
doi = {10.1088/2041-8205/783/1/L21},
url = {https://dx.doi.org/10.1088/2041-8205/783/1/L21},
year = {2014},
month = {feb},
publisher = {The American Astronomical Society},
volume = {783},
number = {1},
pages = {L21},
author = {Sironi, Lorenzo and Spitkovsky, Anatoly},
title = {RELATIVISTIC RECONNECTION: AN EFFICIENT SOURCE OF NON-THERMAL PARTICLES},
journal = {ApJL}
}

@ARTICLE{1978MNRAS.182..147B,
       author = {{Bell}, A.~R.},
        title = "{The acceleration of cosmic rays in shock fronts - I.}",
      journal = {\mnras},
         year = 1978,
        month = jan,
       volume = {182},
        pages = {147-156},
          doi = {10.1093/mnras/182.2.147},
       adsurl = {https://ui.adsabs.harvard.edu/abs/1978MNRAS.182..147B}
}

@ARTICLE{1978ApJ...221L..29B,
       author = {{Blandford}, R.~D. and {Ostriker}, J.~P.},
        title = "{Particle acceleration by astrophysical shocks.}",
      journal = {\apjl},
         year = 1978,
        month = apr,
       volume = {221},
        pages = {L29-L32},
          doi = {10.1086/182658},
       adsurl = {https://ui.adsabs.harvard.edu/abs/1978ApJ...221L..29B}
}

@ARTICLE{Mimica2009,
   author = {{Mimica}, P. and {Aloy}, M.-A. and {Agudo}, I. and {Mart{\'i}}, J.~M. and 
	{G{\'o}mez}, J.~L. and {Miralles}, J.~A.},
    title = "{Spectral Evolution of Superluminal Components in Parsec-Scale Jets}",
  journal = {\apj},
archivePrefix = "arXiv",
   eprint = {0811.1143},
 primaryClass = "astro-ph",
     year = 2009,
    month = apr,
   volume = 696,
   number = 2,
    pages = {1142-1163},
      doi = {10.1088/0004-637X/696/2/1142},
   adsurl = {https://ui.adsabs.harvard.edu/abs/2009ApJ...696.1142M}
}

@ARTICLE{1977AJ.....82..781M,
       author = {{Marscher}, A.~P.},
        title = "{Structure of radio sources with remarkably flat spectra: PKS 0735+178.}",
      journal = {\aj},
         year = 1977,
        month = oct,
       volume = {82},
        pages = {781-784},
          doi = {10.1086/112125},
       adsurl = {https://ui.adsabs.harvard.edu/abs/1977AJ.....82..781M}
}

@article{Agudo2011a,
  author    = {Agudo, Iv{\'a}n and Jorstad, Svetlana G. and Marscher, Alan P. and Larionov, Valeri M. and G{\'o}mez, Jos{\'e} L. and et al.},
  title     = {Location of $\gamma$-Ray Flare Emission in the Jet of the BL Lacertae Object OJ287 More than 14 pc from the Central Engine},
  journal   = {ApJL},
  year      = {2011},
  volume    = {726},
  pages     = {L13},
  doi       = {10.1088/2041-8205/726/1/L13}
}

@article{Agudo2011b,
  author    = {Agudo, Iv{\'a}n and Marscher, Alan P. and Jorstad, Svetlana G. and Larionov, Valeri M. and G{\'o}mez, Jos{\'e} L. and et al.},
  title     = {On the Location of the $\gamma$-Ray Outburst Emission in the BL Lacertae Object AO 0235+164 Through Observations Across the Electromagnetic Spectrum},
  journal   = {ApJL},
  year      = {2011},
  volume    = {735},
  pages     = {L10},
  doi       = {10.1088/2041-8205/735/1/L10}
}

@article{Jorstad2013,
  author    = {Jorstad, S. G. and Marscher, A. P. and Smith, P. S. and Larionov, V. M. and Agudo, I. and et al.},
  title     = {The Connection between $\gamma$-Ray Outbursts and Parsec-scale Jet Activity in 3C 454.3 in 2005--2011},
  journal   = {ApJ},
  year      = {2013},
  volume    = {773},
  pages     = {147},
  doi       = {10.1088/0004-637X/773/2/147}
}

@article{Tavecchio2011,
  author    = {Tavecchio, F. and Becerra-Gonzalez, J. and Ghisellini, G. and Bonnoli, G. and Donato, D. and et al.},
  title     = {On the origin of the $\gamma$-ray emission from the flaring blazar PKS 1222+216},
  journal   = {A\&A},
  year      = {2011},
  volume    = {534},
  pages     = {A86},
  doi       = {10.1051/0004-6361/201117204}
}

@article{Orienti2013,
  author    = {Orienti, M. and Koyama, S. and D'Ammando, F. and Giroletti, M. and Hovatta, T. and et al.},
  title     = {Radio and $\gamma$-ray follow-up of the exceptionally high-activity state of PKS 1510$-$089 in 2011},
  journal   = {MNRAS},
  year      = {2013},
  volume    = {428},
  pages     = {2418--2429},
  doi       = {10.1093/mnras/sts201}
}

@article{Brown2013,
  author    = {Brown, Anthony M.},
  title     = {Locating the $\gamma$-ray emission region of the flat spectrum radio quasar PKS 1510$-$089},
  journal   = {MNRS},
  year      = {2013},
  volume    = {431},
  pages     = {824--835},
  doi       = {10.1093/mnras/stt173}
}

@article{Abdalla2021,
  author    = {{H.E.S.S. Collaboration} and {MAGIC Collaboration}},
  title     = {H.E.S.S. and MAGIC observations of a sudden cessation of a very-high-energy $\gamma$-ray flare in PKS 1510$-$089 in May 2016},
  journal   = {A$\&$A},
  year      = {2021},
  volume    = {648},
  pages     = {A23},
  doi       = {10.1051/0004-6361/202038949}
}

@article{Rani2013,
  author    = {Rani, B. and Krichbaum, T. P. and Marscher, A. P. and Hodgson, J. A. and Fuhrmann, L. and Zensus, J. A.},
  title     = {Constraining the location of rapid $\gamma$-ray flares in the flat spectrum radio quasar 3C~273},
  journal   = {A\&A},
  year      = {2013},
  volume    = {557},
  pages     = {A71},
  doi       = {10.1051/0004-6361/201321440}
}

@article{Ma2025CTA102,
  author    = {Ma, Chen-Li and Deng, Jun-Hao and Jiang, Yun-Guo},
  title     = {Unraveling the Variation Mechanism of CTA 102: A Combined Spectral and Polarization Investigation},
  journal   = {ApJ},
  year      = {2025},
  volume    = {990},
  pages     = {82},
  doi       = {10.3847/1538-4357/adef14}
}

@article{ paraschosa,
	author = {{Paraschos, G. F.} and {Debbrecht, L. C.} and {Kramer, J. A.} and {Traianou, E.} and {Liodakis, I.} and {Krichbaum, T. P.} and {Kim, J.-Y.} and {Janssen, M.} and {Nair, D. G.} and {Savolainen, T.} and {Ros, E.} and {Bach, U.} and {Hodgson, J. A.} and {Lisakov, M.} and {MacDonald, N. R.} and {Zensus, J. A.}},
	title = {Evidence of a toroidal magnetic field in the core of 3C 84},
	DOI= "10.1051/0004-6361/202450218",
	url= "https://doi.org/10.1051/0004-6361/202450218",
	journal = {A$\&$A},
	year = 2024,
	volume = 686,
	pages = "L5",
}

@article{ paraschosb,
	author = {{Paraschos, G. F.} and {Wielgus, M.} and {Benke, P.} and {Mpisketzis, V.} and {Rösch, F.} and {Dasyra, K.} and {Ros, E.} and {Kadler, M.} and {Ojha, R.} and {Edwards, P. G.} and {Hyland, L.} and {Quick, J. F. H.} and {Weston, S.}},
	title = {First very long baseline interferometry detection of Fornax A},
	DOI= "10.1051/0004-6361/202450590",
	url= "https://doi.org/10.1051/0004-6361/202450590",
	journal = {A$\&$A},
	year = 2024,
	volume = 687,
	pages = "L6",
}

@article{Abdollahi_2023,
   title={The Fermi-LAT Lightcurve Repository*},
   volume={265},
   ISSN={1538-4365},
   url={http://dx.doi.org/10.3847/1538-4365/acbb6a},
   DOI={10.3847/1538-4365/acbb6a},
   number={2},
   journal={ApJS},
   publisher={American Astronomical Society},
   author={Abdollahi, S. and Ajello, M. and Baldini, L. and Ballet, J. and Bastieri, D. and Becerra Gonzalez, J. and Bellazzini, R. and Berretta, A. and Bissaldi, E. and Bonino, R. and Brill, A. and Bruel, P. and Burns, E. and Buson, S. and Cameron, R. A. and Caputo, R. and Caraveo, P. A. and Cibrario, N. and Ciprini, S. and Cristarella Orestano, P. and Crnogorcevic, M. and Cutini, S. and D’Ammando, F. and De Gaetano, S. and Digel, S. W. and Di Lalla, N. and Di Venere, L. and Domínguez, A. and Ramazani, V. Fallah and Fegan, S. J. and Ferrara, E. C. and Fiori, A. and Fleischhack, H. and Franckowiak, A. and Fukazawa, Y. and Fusco, P. and Gammaldi, V. and Gargano, F. and Garrappa, S. and Gasbarra, C. and Gasparrini, D. and Giglietto, N. and Giordano, F. and Giroletti, M. and Green, D. and Grenier, I. A. and Guiriec, S. and Gustafsson, M. and Hays, E. and Horan, D. and Hou, X. and Jóhannesson, G. and Kerr, M. and Kocevski, D. and Kuss, M. and Latronico, L. and Li, J. and Liodakis, I. and Longo, F. and Loparco, F. and Lorusso, L. and Lott, B. and Lovellette, M. N. and Lubrano, P. and Maldera, S. and Manfreda, A. and Martí-Devesa, G. and Mazziotta, M. N. and Mereu, I. and Meyer, M. and Michelson, P. F. and Mizuno, T. and Monzani, M. E. and Morselli, A. and Moskalenko, I. V. and Negro, M. and Omodei, N. and Orlando, E. and Ormes, J. F. and Paneque, D. and Panzarini, G. and Perkins, J. S. and Persic, M. and Pesce-Rollins, M. and Pillera, R. and Porter, T. A. and Principe, G. and Racusin, J. L. and Rainò, S. and Rando, R. and Rani, B. and Razzano, M. and Razzaque, S. and Reimer, A. and Reimer, O. and Sánchez-Conde, M. and Parkinson, P. M. Saz and Scargle, Jeff and Scotton, L. and Serini, D. and Sgrò, C. and Siskind, E. J. and Spandre, G. and Spinelli, P. and Suson, D. J. and Tajima, H. and Thompson, D. J. and Torres, D. F. and Valverde, J. and Venters, T. and Wadiasingh, Z. and Wagner, S. and Wood, K.},
   year={2023},
   month=mar, pages={31} }

@ARTICLE{2019AA...630A..56P,
       author = {{Pati{\~n}o-{\'A}lvarez}, V.~M. and {Dzib}, S.~A. and {Lobanov}, A. and {Chavushyan}, V.},
        title = "{Is there a non-stationary {\ensuremath{\gamma}}-ray emission zone 42 pc from the 3C 279 core?}",
      journal = {\aap},
         year = 2019,
        month = oct,
       volume = {630},
          eid = {A56},
        pages = {A56},
          doi = {10.1051/0004-6361/201834401},
archivePrefix = {arXiv},
       eprint = {1907.08314},
 primaryClass = {astro-ph.HE},
       adsurl = {https://ui.adsabs.harvard.edu/abs/2019A&A...630A..56P}
}

@ARTICLE{1994ApJ...421..153S,
       author = {{Sikora}, Marek and {Begelman}, Mitchell C. and {Rees}, Martin J.},
        title = "{Comptonization of Diffuse Ambient Radiation by a Relativistic Jet: The Source of Gamma Rays from Blazars?}",
      journal = {\apj},
         year = 1994,
        month = jan,
       volume = {421},
        pages = {153},
          doi = {10.1086/173633},
       adsurl = {https://ui.adsabs.harvard.edu/abs/1994ApJ...421..153S}
}

\clearpage

\begin{appendix}

\section{Calibration of the instrumental polarization}
\subsection{Polarization leakage and D-terms}
\label{ap:pol_cal}

The polarization calibration was performed in AIPS using the LPCAL task, which estimates the D-terms for each intermediate frequency (IF) independently. We used \object{TXS~2013+370} itself as the polarization calibrator, providing LPCAL with a Stokes I source model derived from our imaging process. LPCAL employs a linear fitting algorithm that requires polarization leakage $<5\%$ and good parallactic angle coverage for accurate results. The VLBA antennas achieved large parallactic angle coverage ($\geq 100\degr$, reaching $\sim 200\degr$), resulting in leakages $<5\%$. However, Effelsberg showed limited coverage ($\sim 15\degr$), leading to $\sim 15\%$ leakage at 22\,GHz and unreliable values at 43\,GHz. Comparison of calibration results with and without Effelsberg showed that this antenna provided no additional information for the new jet component, only confirming the core EVPA already determined by VLBA data. Therefore, we excluded Effelsberg from polarization analysis while retaining its data for total intensity imaging.

The computed D-terms for VLBA antennas showed consistent values across all IFs, with dispersions of $\sim 2.4\%$ in amplitude and $\sim 77.7\degr$ in phase for 22~GHz, $\sim 1.7\%$ and $\sim 94.2\degr$ for 43~GHz, and $\sim 6.9\%$ and $\sim 93.2\degr$ for 86~GHz. For each antenna, we calculated median values across the IFs as representative D-terms.

Absolute EVPA calibration was performed using quasi-simultaneous single-dish observations from the POLAMI monitoring program\footnote{See \cite{10.1093/mnras/stx2435} and http://polami.iaa.es}. For 22 and 43\,GHz, we used polarization angles measured on 9 February 2021 ($109\degr \pm 3.6\degr$ at 13~mm) and 10 February 2021 ($113.9\degr \pm 0.5\degr$ at 7~mm), respectively. For 86\,GHz, we used Cubic Spline interpolation between the polarization angle and time, from historical POLAMI data, to obtain via extrapolation $51.9\degr \pm 7.9\degr$ for 11 February 2021. 

We also estimated the polarization parameters and their uncertainties for all data sets, following \cite{Gomez_2022}, with the results presented in Table \ref{table:obs_par}. The parameters were computed using data above a threshold of $5~\times$ rms, where rms is an estimation of the map thermal noise.

\label{appendix:Dterms}

\begin{figure}[h!]
\centering
\begin{subfigure}{\columnwidth}
\includegraphics[width=.9\linewidth]{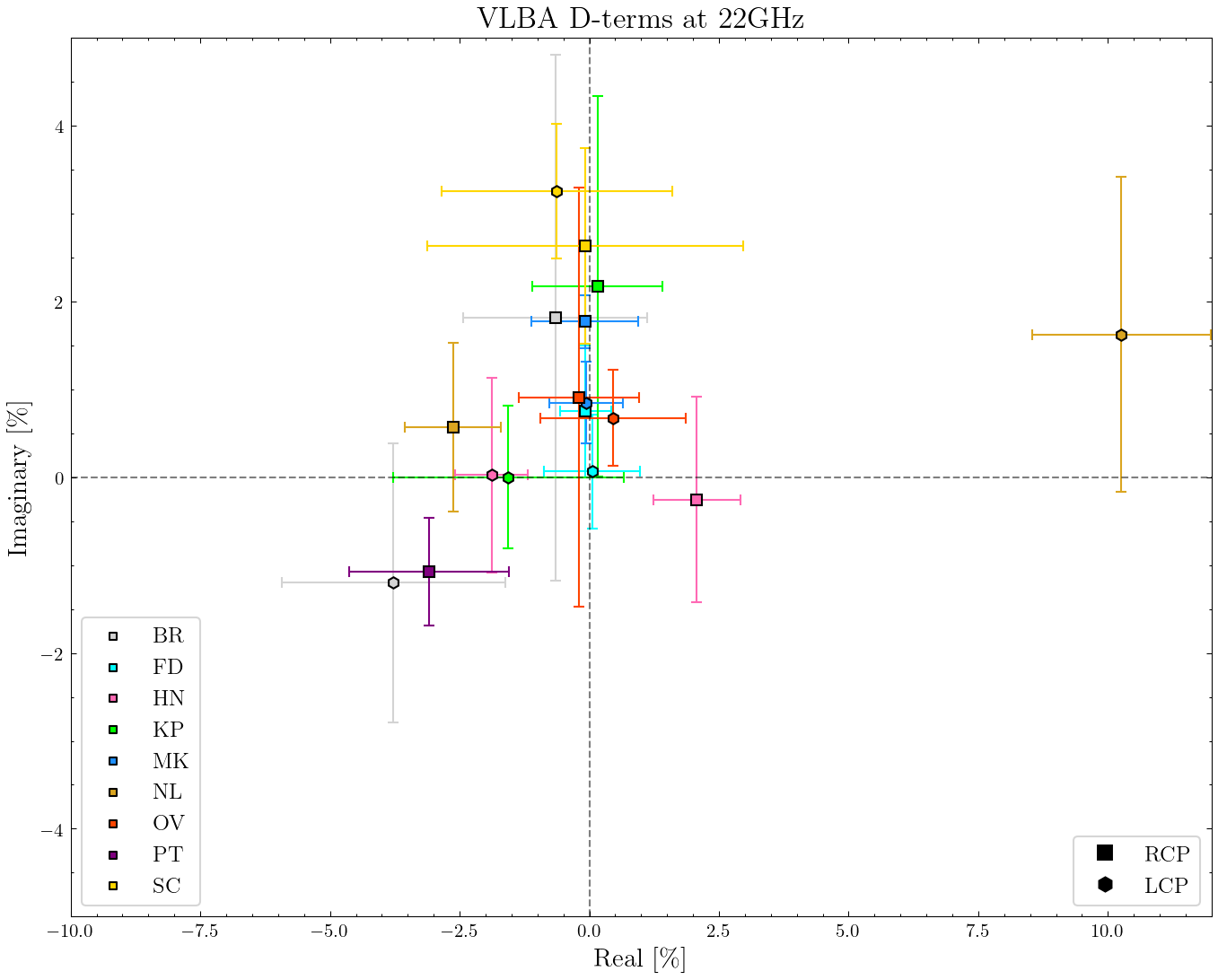}
\end{subfigure}

\begin{subfigure}{\columnwidth}
\includegraphics[width=.9\linewidth]{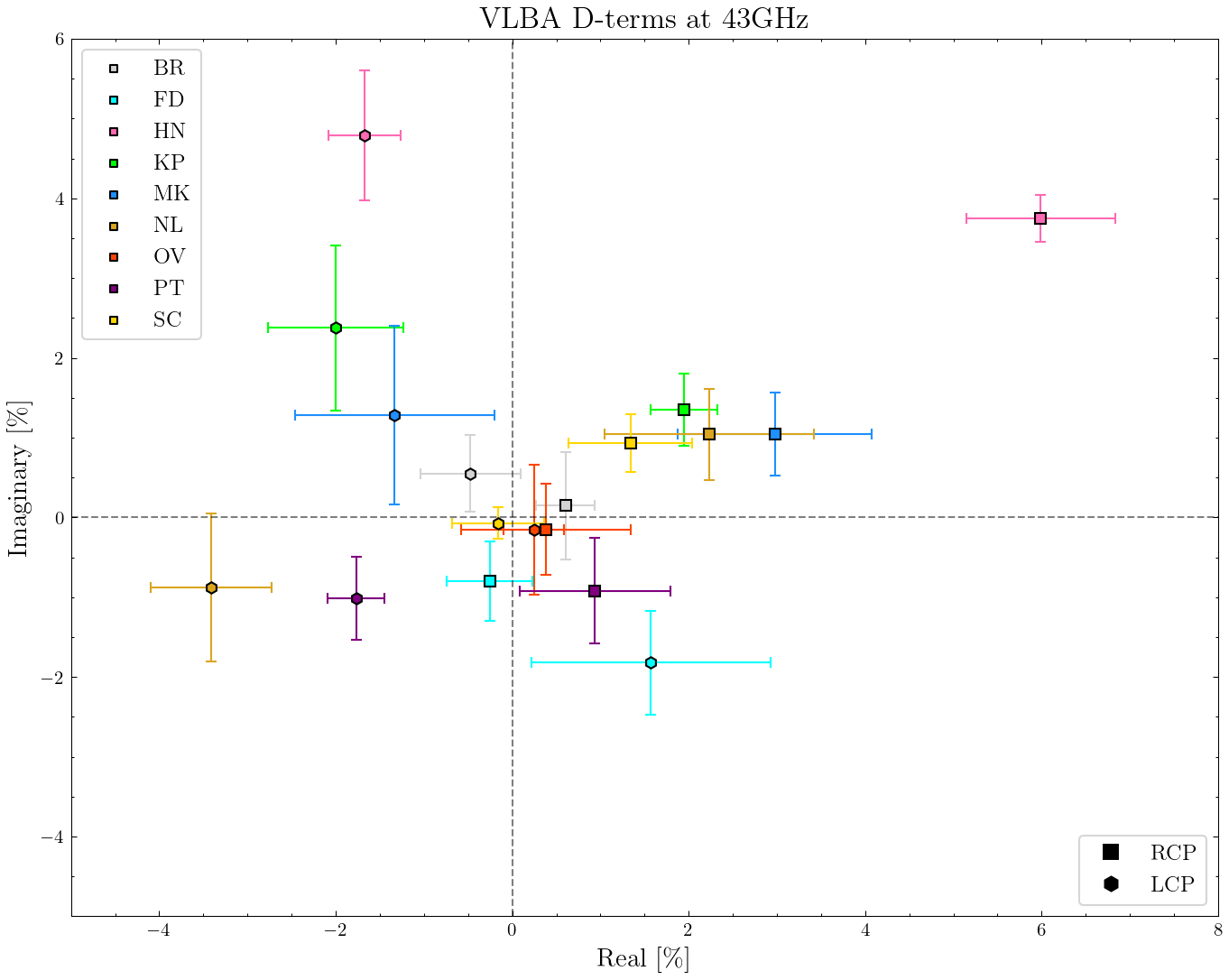}
\end{subfigure}

\begin{subfigure}{\columnwidth}
\includegraphics[width=.9\linewidth]{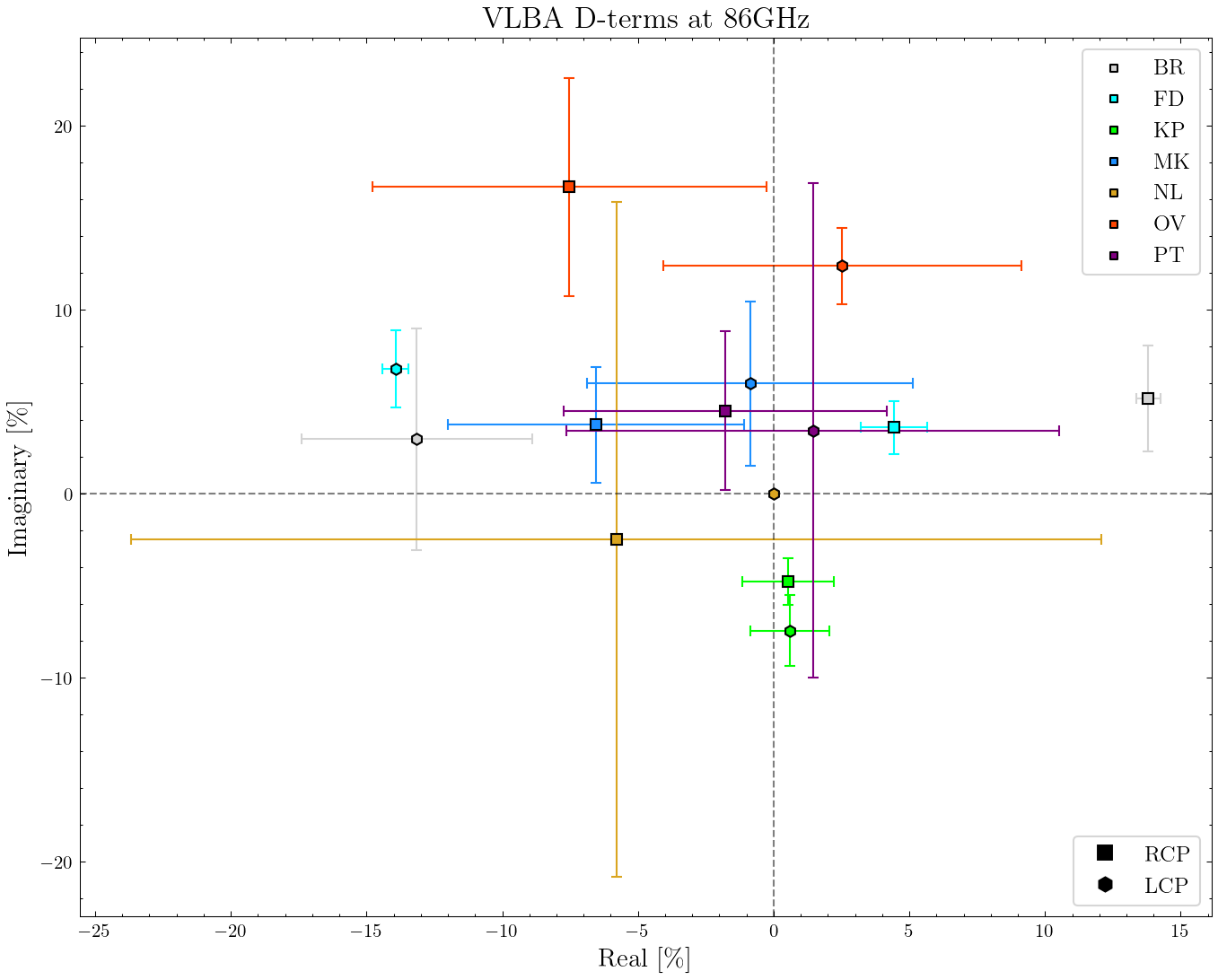}
\end{subfigure}

\caption{D-terms for the VLBA observations at 22, 43, and 86~GHz for RCP and LCP. Plotted values correspond to the medians of 4 IFs, with errors computed from the standard deviation across all IFs.}
\label{fig:dterms}
\end{figure}

\end{appendix}

\end{document}